\documentclass[aps]{revtex4}

\begin{document}

\title{NONPERTURBATIVE DESCRIPTION OF THE MASS AND CHARGE RENORMALIZATION IN QUANTUM ELECTRODYNAMICS }

\author{I D Feranchuk}

\address{Belarussian State University,4, F.Skariny Av., 220050 Minsk, Republic of Belarus \\
E-mail: fer@open.by\\
Tel.: (375 172) 277 617; Fax: (375 172) 202 353}

\begin{abstract}

\noindent \textbf{Abstract}\\
In this paper the nonperturbative analysis of the spectrum for one-particle excitations of the
electron-positron field (EPF) is considered in the paper. A standard form of the quantum electrodynamics (QED)
is used but the charge of the "bare" electron $e_0$ is supposed to be of a large value ($e_0 \gg 1$). It is
shown that in this case  the quasi-particle can be formed with a non-zero averaged value of the scalar component
of the electromagnetic field (EMF). Self-consistent equations for the distribution of charge density in the
"physical" electron (positron) are derived. A variational solution of these equations is obtained and it defines
the finite renormalization of the charge and mass of the electron (positron). It is found that the coupling
constant $\alpha_0$ between EPF and EMF and mass  $m_0$ of the "bare" electron can be connected with the
observed values of the fine structure constant  $\alpha$ and the mass of the "physical" electron $m$ as follows
($\hbar = c = 1$):

$$
\alpha_0 \sim \frac{1}{\alpha}; \qquad m_0 \sim \frac{m}{\alpha}.
$$

It is also shown that although the non-renormalized QED corresponds to the strong coupling between EPF and EMF,
the interaction between "physical" electron (positron) with EMF is defined by the observed value of the coupling
constant $\alpha$. It is proved that the translational motion of the "physical" particle is separated from its
internal degrees of freedom. As a result the dependence of the one-particle excitation energy on its total
momentum is defined by the formula $E(\vec P) = \sqrt{P^2 + m^2}$ and corresponds to the relativistic spectrum
of a free particle with the observed value of mass m. Regularization of the terms of a series of the
perturbation theory by  $\alpha$  is due to the form-factor of the "physical" electron (positron) and
corresponds to the cut-off momentum   $k_{max} \simeq m_0$.
\end{abstract}

\maketitle

\newpage


Short Title: NONPERTURBATIVE QUANTUM ELECTRODYNAMICS

\maketitle

\newpage

\section{Introduction}

Quantum electrodynamics (QED) is one of the basic elements of modern physics foundation. It allows to calculate
the observed characteristics of fundamental electromagnetic processes  with a unique accuracy. This means
undoubtedly that the basic equations of QED correctly describe possible states of the electromagnetic (EMF) and
electron-positron (EPF) fields. It is also well known that the problem of describing one-particle excitations of
EPF can be formulated consistently in the framework of QED. The characteristics of these quasi-particles should
determine the renormalization of the electron (positron) mass and charge because of their interaction with EMF.
But the real solution of this problem in terms of the perturbation theory leads to some infinite values. An
important property of QED is its renormalizability. It means that a special procedure can be developed which
permits  to exchange formally infinite parameters for their observed values when calculating the characteristics
of  electromagnetic processes. But at the same time Dirac wrote that "the calculation rules of QED are badly
adjusted with the logical foundations of quantum mechanics and they can't be considered as the satisfactory
solution of the difficulties" \cite{Dirac}($\S 81$).

It's worth reminding that these rules are based either on  some way of regularization for diverged integrals
(for example, \cite{Akhiezer}) or on adding counter-terms to Lagrangian of the system \cite{Bogoliubov}. As a
result all observed characteristics are finite values and only unobserved values of the charge $e_0$ and mass
$m_0$ of the "bare" electron (positron) remain indeterminate because they include the diverged integrals.
Recently a successive way for "finite QED" was formulated \cite{Scharf95}, \cite{Scharf01}. In this case the
perturbation theory series is calculated strictly for elements of the scattering matrix $(S-matrix)$. The
regularization of integrals is realized by means of accurately calculating the singular functions and the
problem of describing the "physical" particle internal structure doesn't arise at all.

Thus, from the "pragmatic" point of view the existing form of  QED is a perfect technique for describing
observed electromagnetic phenomena. Nevertheless, this approach is not completely accurate because "it is simply
a way to sweep the difficulties under the rug" as Feynman wrote
 \cite{Feynman}.

What can stimulate  searching for the solution of the problem of calculating the "bare" electron characteristics
$e_0$ and $m_0$ in spite the fact that these values are unobservable and are not strictly included into
experimentally measured  physical values? First of all the possibility of their successive calculation is
connected with the question whether the system of Maxwell and Dirac equations is closed and self-consistent when
considering the processes involved with the interaction of electrons, positrons and photons? It is important to
stress that we mean the internal consistency of QED only as the mathematical model for quantum field system. It
can't be considered closed from the physical point of view because of its interaction with other quantum fields.
In that sense QED in the existing form of renormalization is an unclosed theory because it includes an
additional, external dimensional parameter which enables the regularization of integrals.

Besides, the analysis of the "physical" electron structure could clarify the question about the continuity of
space. The matter is that  fundamental length was introduced in some papers as an additional parameter of the
theory in order to realize the finite renormalization in QED.

There are a lot of papers where the singularity-free electrodynamics is considered on the basis of a classical
model for electron and vacuum polarization (see, for example, the more recent one \cite{Blinder} and references
therein). But it is very important to consider the renormalization problem in the framework of the "logical
principles of quantum mechanics" and relativistic invariance in order to develop some new nonperturbative
approach in the field theory in application to the real physical system with completely defined Hamiltonian. At
present nonperturbative methods are mainly studied for quite abstract quantum field models with a strong
interaction (for example, \cite{Kleinert}).

An essential role in motivating our work is played by the obvious contradiction between the small value of the
coupling constant and its infinite calculated value showing a logical inconsistency of the  perturbation theory
as the main method used in QED.

It is well known  that the physical picture of the charge renormalization is mainly defined by the effect of
polarization of  electron-positron vacuum \cite{Lifshitz}. However, as the authors \cite{Lifshitz}(\S 128)
noticed, it is "logically inconsistent" to analyze the connection between the charges $e_0$ and $e$ in terms of
relations based on the perturbation theory. So, in order to consider the renormalization problem we should
reveal both the reason of divergence of the perturbation series and find  the way for nonperturbative
description of the system.

In the present paper we suggest a possible variant for a nonperturbative investigation of the physical picture
corresponding to the charge screening conditioned by the polarization of  electron-positron vacuum. From the
very beginning the initial charge is supposed to  be larger than 1 ( $e_0 \gg 1$). It means that the
nonrenormalized QED should be considered as the system of interacting fields with a strong coupling. It was
first shown by Pekar \cite{Pekar} and then analyzed more rigorously by Bogoluibov and Tyablikov
\cite{Bogol},\cite{Tyablikov}, that the main physical result of the strong coupling between the particle and the
quantum field is defined by the self-localized ("polaronic") state of the particle. This state is due to the
classical components selected from the quantum field because of its interaction with the particle. Later a lot
of different forms of the quasi-particles arising as a result of a "polaronic" effect were described for many
concrete physical systems. Another way for nonperturbative calculations in QED was considered recently in the
paper \cite{Rochev}.

In this paper it is shown that the self-localized quasi-particle excitation of the EPF corresponds to the
solution of QED equations for the case of  $e_0 \gg 1$. The internal structure of such excitation is determined
by a spatial charge distribution of EPF coupled with a classical scalar component of EMF separated by the same
charge density. We associate such excitation with the "physical" electron (positron). It allows us to establish
the analytical relation between QED initial parameters
 $e_0, m_0$ with their observed values $e, m$ and to justify the self-consistency of the original supposition.

When describing the observed electromagnetic processes the relativistic invariance and renormalizability are the
most important properties of QED confirmed by a lot of precision experiments. Therefore the self-consistent
dynamic form-factors, which we have introduced for the internal structure of the "physical" electron, should be
relativistically invariant and reveal themselves in the observed physical characteristics only in the form of
the renormalized charge and mass of the electron at least up to very high energies of interacting particles. The
analysis of this problem is considered in the second part of the paper. Our purpose is to substantiate that the
supposition about the strong coupling between EPF and EMF doesn't logically contradict  the standard
perturbation theory in the renormalized QED.

The paper is organized as follows. In Section 2 we consider the qualitative analysis of the physical picture of
the electron-positron vacuum polarization when the quasi-particle excitation is formed. The use of quite simple
scale estimation permitted us to find the characteristic correlations between the observed charge and mass of
the electron and the initial parameters of the theory. Besides, the mathematical reason for the divergence of
the perturbation series in QED is discussed.

In Section 3 the method for nonperturbative calculation of the characteristics of self-localized one-particle
excitation of the electron-positron field is developed in the framework of QED. The system of self-consistent
equations for describing the charge distribution and the potential of the electrostatic field coupled with it is
derived from the zero approximation of the approach under consideration. The total momentum of the excitation is
supposed to be zero in this case. The variational solution of these equations is found and more precise
relations between the charge and mass of "bare" electron and the corresponding values for "physical" electron
are calculated.

The dependence of the quasi-particle excitation energy $E(\vec P)$ on its total momentum is found in Section 4.
As is known such dependence is rather complicated in the "polaron" problem and it is very difficult to find
analytically its general form  because the variables describing the internal dynamics of the system are not
separated from the tranlational variables \cite{Gross}. Therefore the calculation of $E(\vec P)$ for QED is a
very important verification for the nonperturbative  approach in question when describing the observed
properties of the "physical" electron (positron). It is shown in this section that nonperturbative QED satisfies
all necessary conditions: the internal degrees of freedom are separated from the translational motion of the
"physical" electron (positron) and its dispersion law $E(\vec P)$ corresponds to the standard dynamics of the
relativistic free particle.

One of the main purposes of our paper is differs greatly from most of the theoretical works where some results
for experimental observation are predicted. On the contrary, we are going to prove that the internal structure
of the "physical" electron considered in our paper doesn't reveal in its observable characteristics. Therefore,
in Section 5 the rules for transition from the initial theory with the strong coupling to  the renormalized QED
are justified. It is shown that the interaction between the "physical" electron and the transversal
electromagnetic field is defined by the observed electron charge. It means that the characteristics of real
electromagnetic processes can be calculated using the standard perturbation theory, regularization of the
diverged integrals in QED being defined by the form-factor of the charge distribution in the "physical" electron
(positron).

\section{Qualitative analysis}

Let us first discuss the qualitative character of the processes which can determine the formation of the
"physical" electron (PE). We use Dirac's interpretation for the ground state of the vacuum electron-positron
field (VEPF) \cite{Dirac}. As is known (for example, \cite{Heitler}),  such representation for the solutions of
Dirac equation with negative energies is not rigorous enough but allows one to interpret intuitively the
qualitative picture of the effects associated with the creation of pairs and polarization of VEPF in strong
electromagnetic fields.

It should be noted, that in this picture the ground state of VEPF is taken to be  the energy zone completely
filled by the "bare" electrons (BE) with the charge  $(-e_0), \quad e_0 > 0$ up to the negative maximal energy
$(- m_0)$. The empty states of the continuous spectrum are separated from the border of the vacuum zone by the
energy gap $\Delta_0 = 2 m_0$,  with $m_0$ being the mass of BE.

We will proceed  from the above mentioned statement that QED is the closed model for two interacting fields in
the sense that all the problems of renormalization in this theory are considered to be  the result of some
mathematical difficulties. It means that they should be overcome without including the interaction with other
fields and without any changes in the QED Hamiltonian for regularization. Thus, all further constructions are
based on the assumption that the values $m_0$ and $e_0$ are the only parameters of the theory, provided we use
the system of units with $\hbar = c = 1$.

If the Dirac representation for the energy structure of VEPF is used, the one-particle excitation can be
regarded as the appearance in the free energy zone of the "bare" electron localized near some point of  space.
Generally speaking, the assumption about local excitation breaks the translational symmetry of the system. It
will be shown below (Section 4) that taking into account this symmetry one can find the dependence of the
excitation energy on the total momentum and this energy spectrum  proves to be relativistically invariant.
Actually, in our qualitative consideration we discuss the quasi-classical picture for the formation of the state
corresponding to  "physical" electron (positron) at rest. To some extent this analysis is analogous to the
estimation of the binding energy of the self-localized state of the electron in the ionic crystal which was
considered in the first Pekar's work on the "polaron" theory \cite{Pekar}.

The electrostatic interaction between the excited BE and the electrons from a lower zone determines the
polarization of VEPF which leads to some redistribution of the electrons in the filled zone and localized near
the excited BE. In fact, it means that a surplus positive charge appears in this local domain and this charge
can be described by some distribution of the "bare" positrons (BP). Neglecting the spin effects, the interaction
between different charges in this stationary excited state can be defined mainly by the local self-consistent
potential (LSP) of attraction $\varphi(r)$ which corresponds to the classical part of the scalar component of
EMF. It is supposed that the coordinate  $\vec{r}$ is counted off from the center of the mass of the
quasi-particle.

Let  the amplitude of LSP reaches its maximal value $\varphi_m$ in some point. Then the energy levels for the
states of BE and BP near this point will be shifted down and up respectively along the energy scale by the
values being proportional to the values $e_0 |\varphi_m|$. As QED is supposed to be the closed theory these
shifts should be of the following form, taking into account following from the dimensional arguments:

\begin{eqnarray}
\label{Eqn1} \Delta_{\pm} = m_0 f_{\pm}(e_0),
\end{eqnarray}

and the functions $f_{\pm}(e_0)$ should be of the same form because of the charge symmetry.

As a result, the real value for the energy gap between the energy of excitation and the top of the VEPF zone is

$$
\Delta = \Delta_+ - \Delta_-
$$
\noindent and it depends on the binding energy of the excited BE with the cloud of BP appearing due to the
polarization of VEPF.

The binding energy depending on the initial charge  $e_0$, it is possible to simulate the qualitative behavior
of the dimensionless functions $f_{\pm}$ by the curves represented by the dotted lines in  Figure 1. Let's
denote by $e^*_0$ that value of the BE charge at which  the energy level of the one-particle excitation
intersects locally the filled zone of VEPF. Evidently, the real value of  $e_0$ is unknown a priori and we
consider two "scenarios" of the renormalization which are qualitatively different.

In the case of  $e_0 < e^*_0$, the sequence of the excited energy levels in the real system adiabatically
corresponds to the structure of the energy zones in the ideal system. As a rule, it is one of the necessary
conditions for the convergence of the series of the perturbation theory by the coupling constant \cite{Mors}.

However, as the estimation given below shows,  the real QED is apparently defined by another variant, i.e.  $e_0
> e^*_0$, corresponding to the strong coupling limit. In this case the quasi-intersection of the one-particle
excited levels with the filled VEPF zone arises. As a result a new structure of the energy zones is to be formed
upon including the interactions for "physical" particles. Fig.1 shows this scheme with  solid lines. An
analogous scheme of the level "reconstruction" is considered for the effect of the pair creation in the
homogeneous electric field \cite{electric} or in the Coulomb field of heavy atoms  \cite{kulon}. Similar
behavior of the energy levels is observed in solid state physics when describing the "resonant" zones (for
example, \cite{Zimann}). So, in the case of $e_0 > e^*_0$ one should expect the non-adiabatic correspondence
between the levels of the "physical" electron-positron field and the basic set of the system levels without any
interaction.

The qualitative analysis of the possible behavior of the excited energy levels in QED allows us to establish two
essential circumstances. Firstly, if the real value of $e_0$ is indeed in the range  $e_0
> e^*_0$, then the perturbation series for describing FEL structure, based on the set of states of
the ideal system in the absence of interaction should be a diverged one because of the non-adiabatic change of
the states when taking into account the interaction between EPF and EMF.

Secondly, as Fig.1 shows, the state corresponding to the top level of the filled (that is positronic) zone for
"physical" EPF  is actually the continuation of the bottom level of the empty (that is electronic) zone of VEPF
if the level classification for non-interacting fields is used. It means that the quasi-particle excitation of
VEPF in the "bare" electron form results in the state that should be considered as the "physical" positron (!).
This statement has also a qualitative explanation. When the zones are locally overlapped due to the VEPF
polarization, some part of BE from the lower zone passes to free levels of the upper zone. Then they extend to
infinity because of the repulsion with the initially negative fluctuation of VEPF. As the result, the screening
cloud with the surplus positive charge is formed near the initial BE. Certainly, this intuitive arguments will
be verified by analytical calculations in the following sections.

Here we will consider a very simple and approximate quantitative analysis of the above qualitative picture of
the renormalization. It is evident that the accurate description of the quasi-particle excitation of VEPF
requires the solution of the related system of Dirac's equations for the electron-positron field and Maxwell's
equations for the electromagnetic field (for example, \cite{Akhiezer}) that will be considered below ($\S 3$).
At the same time it is known that in the systems with Coulomb interaction there are no other dimensional values
except the electron mass. In many cases it allows to find quite simple scaling estimations which define the
physical characteristics of the system with the accuracy up to some numerical coefficients. So, in this section
we use the quasi-classical scale analysis in order to define characteristic features of the "physical" electron
or positron depending on the initial parameters of QED, taking into account the uncertainty relation but without
considering the spin characteristics.

In order to clarify the essence of this approach let's use it for estimating the energy of the ground state of
the relativistic hydrogen-like atom with the nucleus charge Z. A similar estimation for a non-relativistic atom
is well known, but it seems that it has not been considered for the relativistic case yet. If we consider the
normalized wave function the quasi-classical formula for the relativistic energy of the hydrogen-like atom can
be written as

\begin{eqnarray}
\label{Eqn2} E_H \simeq \sqrt{(\delta p)^2 + m^2} - \frac{Z\alpha}{a},
\end{eqnarray}
\noindent where the value $a$ is the characteristic linear size of the domain where the electron is localized.
It is related to the characteristic fluctuation of its momentum  $\delta p$ by the uncertainty relation

\begin{eqnarray}
\label{Eqn3} u \equiv \delta p \simeq \frac{1}{a}.
\end{eqnarray}

Here the fine structure constant is introduced
$$
\alpha = \frac{e^2}{4\pi}.
$$

Taking the value u as the variational parameter it is possible to find the equation for it from the energy
minimization. As the result the following estimation for the energy can be found

\begin{eqnarray}
\label{Eqn4} \frac{u}{\sqrt{u^2 + m^2}} - Z\alpha = 0; \qquad u =
\frac{Z m \alpha}{\sqrt{1 - (Z\alpha)^2}}; \nonumber\\
E_H = m \sqrt{1 - (Z\alpha)^2}.
\end{eqnarray}

It turns out that the quasi-classical estimation leads to the exact value for the ground state energy of the
relativistic hydrogen atom as it does in the non-relativistic case. This coincidence may be accidental, but it
should be remembered that it usually testifies the existence of closed classical trajectories in the system. The
most essential point for us is that this scale estimation correctly defines that critical value of the nucleus
charge at which Dirac energy gap disappears and the creation of the electron-positron pairs becomes possible
\cite{kulon}.

Let us apply a similar estimation to considering the problem of the formation of the quasi-particle excitation.
In this case an additional degree of freedom characterizes the physical structure of the quasi-particle together
with the values of u and a. So we should introduce one more parameter for it. In fact, if the initial
fluctuation of VEPF corresponds to BE its integral charge should be equal to the value $(-e_0)$. At the same
time the integral screening charge $q_+$ induced by the vacuum polarization may not be coincide with $e_0$ for a
general case. The difference between these values defines the observed quasi-particle charge $e$ which
corresponds to "physical" electron. Thus we introduce the additional variational dimensionless parameter $x$ as
the value defining the probability $P_+$ of finding the "bare" positron in the polarization cloud around the
"bare" electron. As we consider the excitation normalized by one particle, then by  definition

\begin{eqnarray}
\label{Eqn4a} P_+ = \frac{x}{1 + x} ; \qquad P_- = \frac{1}{x + 1},
\end{eqnarray}
\noindent that is the value $x=0$ corresponds to the state with pure BE.

Then the charge balance in the quasi-particle excitation is

\begin{eqnarray}
\label{Eqn5} q_+ = \frac{x}{1 + x} e_0; \qquad e = - (q_- - q_+) = \frac{x - 1}{x + 1} e_0.
\end{eqnarray}

Note that we don't fix the sign of the observed charge $e$ relative to the charge of the excited BE (- $e_0$).

It is evident, that in the case of the standard perturbation theory the one-electron excitation of EPF is
defined by the parameter  $x = 0$. Let us indicate, however, that even the simple scale estimation shows that
the minimal energy of the one-particle excitation is achieved with a non-trivial value of the parameter x.

Suppose that the momentum fluctuations are equal for the electron and positron states forming the self-localized
excitation. Then the quasi-classical estimation for the energy of the "physical" excitation can be written as
follows if one takes into account the physical meaning of the parameter x

\begin{eqnarray}
\label{Eqn6} \sqrt{u^2 + m^2_0}\frac{1 - x}{1+x} - e_0\frac{1 - x}{1+x}<\varphi(r)> - E = 0.
\end{eqnarray}

Now we should estimate the average value of the self-consistent potential by means of the introduced parameters.
The potential itself depends on the local distribution of the positive and negative charge densities  $\pm
e_0\rho_{\pm}(r)$ in the quasi-particle and is defined by the Poisson equation which has the following form in
the chosen system of units

\begin{eqnarray}
\label{Eqn7} \varphi(r) = \frac{e_0}{4\pi} \int d \vec{r}'\frac{[\rho_+(r') - \rho_-(r')]}{|\vec{r} -
\vec{r}'|}.
\end{eqnarray}

If no additional parameters for the local charge distribution are used, then the scale consideration leads to
the following integral estimation for the averaged potential

\begin{eqnarray}
\label{Eqn8} <\varphi(r)> \simeq \frac{e_0}{4\pi}\frac{(x - 1)}{(x + 1)a} = \frac{e_0}{4\pi}\frac{(x -
1)}{(x+1)} u.
\end{eqnarray}

Substituting the estimation (\ref{Eqn8}) into Equation (\ref{Eqn6}) we can find the closed variational formula
for determining the unknown parameters $x, u$ and energy $E$

\begin{eqnarray}
\label{Eqn9} \sqrt{u^2 + m^2_0}(1 - x^2) + \alpha_0(1 - x)^2 u - E(1 + x)^2 = 0;
\nonumber\\
\alpha_0 = \frac{e^2_0}{4\pi}.
\end{eqnarray}

Varying on the parameters leads to

\begin{eqnarray}
\label{Eqn10} \frac{u}{\sqrt{u^2 + m^2_0}}(1 - x^2) + \alpha_0(1 - x)^2 = 0.
\end{eqnarray}

\begin{eqnarray}
\label{Eqn11} - 2 x\sqrt{u^2 + m^2_0} + 2 \alpha_0 (x - 1) u - 2 E (1+x) = 0.
\end{eqnarray}

The solution of Equation (\ref{Eqn10}) exists only if $x > 1$ and  corresponds to the above given
renormalization variant when $e_0 > e^*_0$, so one can find

\begin{eqnarray}
\label{Eqn12} u = \frac{m_0 \alpha_0 t}{\sqrt{1 - \alpha^2_0 t^2}}; \quad t = \frac{x - 1} {x + 1};
\nonumber\\
\sqrt{u^2 + m^2_0} = \frac{m_0}{\sqrt{1 - \alpha^2_0 t^2}}.
\end{eqnarray}

Substituting this solution into Eqs.  (\ref{Eqn11}) and (\ref{Eqn9}) leads to

\begin{eqnarray}
\label{Eqn13} - \frac{x}{1+x} +  \alpha^2_0 t^2 = \epsilon \sqrt{1 - \alpha^2_0 t^2};
\nonumber\\
- t + \alpha^2_0 t^3 = \epsilon \sqrt{1 - \alpha^2_0 t^2};
\nonumber\\
E = m_0 \epsilon,
\end{eqnarray}
\noindent with $\epsilon$  being the dimensionless parameter defining the self-energy of the quasi-particle.

Excluding  $\epsilon$ from the last two equations leads to the equation for the screening parameter  $x$:

\begin{eqnarray}
\label{Eqn14} 2 \alpha^2_0 (x - 1)^2 = (x + 1)^2.
\end{eqnarray}

This equation relates $x$ to the initial charge $e_0$ which, in its turn, can be expressed by the observed
charge e of the quasi-particle by means of the normalized condition (\ref{Eqn5}). The explicit form for the
values $x$ and $\epsilon$ is:

\begin{eqnarray}
\label{Eqn15} t = \frac{1}{\alpha_0 \sqrt{2}}; \quad x = \frac{\alpha_0 \sqrt{2} + 1} {\alpha_0 \sqrt{2} - 1};
\quad \sqrt{1 - \alpha^2_0 t^2} = \frac{1}{\sqrt{2}};
\nonumber\\
E = - m_0 t \sqrt{1 - \alpha^2_0 t^2} = - \frac{m_0}{2\alpha_0}.
\end{eqnarray}

Substituting these equations into the normalized condition (\ref{Eqn5}) shows that according to the qualitative
analysis the observed charge of the quasi-particle proves to be positive ($e > 0$), i.e. it corresponds to
"physical" positron . This result is also consistent with the negative value of the self energy of the
quasi-particle that defines the top of the "physical" positron zone (-m). Its absolute value should be equal to
the observed mass of the "physical" positron (see Fig.1). Thus, in the framework of the accuracy of the
considered quasi-classical estimation the connection between renormalized and initial parameters of QED is
defined by the formulas clearly indicating the nonperturbative character of the above calculation

\begin{eqnarray}
\label{Eqn16} e_0 \frac{1}{\alpha_0\sqrt{2}} = e; \qquad \alpha_0 = \frac{1}{2 \alpha};
\nonumber\\
m = \frac{m_0}{2\alpha_0}; \qquad  m_0 = \frac{m}{ \alpha}.
\end{eqnarray}

To describe the characteristics of the "physical" electron one has to repeat all the calculations considering
"bare" positron as the initial fluctuation of VEPF, not changing the values of the renormalization parameters
because of the charge symmetry. The numerical values  of the initial parameters of QED are of some interest,
too:

\begin{eqnarray}
\label{Eqn17}
\alpha_0 = \frac{1}{2\alpha} \simeq 68.5;\nonumber\\
m_0 = \frac{m}{\alpha} \simeq 70 MeV.
\end{eqnarray}

\section{Variational equations for one-particle excitation in QED}

As  mentioned above, the Dirac interpretation of the energy levels of the vacuum electron-positron field is
quite qualitative and the estimation obtained on its basis in Section 2 is useful mainly for visualizing the
formation of the quasi-particle excitation, corresponding to the "physical" particle. Therefore it is important
to answer the question: whether there exists the configuration of the interacting fields which leads to the
analogous structure of one-particle excitation in the framework of the accurate formulation of QED?

Let's use the hamiltonian QED in the Coulomb gauge \cite{Heitler}. This representation is most convenient for us
because it extracts the electrostatic field which, by definition, makes the main contribution to the vacuum
polarization in the limit of strong coupling:

\begin{eqnarray}
\label{Eqn18} \hat H = \int d \vec{r}\{ \hat \psi^* (\vec{r}) [ \vec \alpha (\vec{p} + e_0 \hat{ \vec{ A} }
(\vec{r})) + \beta m_0] \hat \psi (\vec{r}) +  e_0\hat \varphi (\vec{r}) \hat \rho (\vec{r}) - \frac{1}{2} (
\vec{\nabla}\hat \varphi
(\vec{r}))^2\} + \sum_{\vec k \lambda} \omega(\vec k) \hat n_{\vec k \lambda};\nonumber\\
\hat \rho (\vec{r}) = \frac{1}{2} [\hat \psi^* (\vec{r})\hat \psi (\vec{r}) - \hat \psi (\vec{r})\hat \psi^*
(\vec{r})].
\end{eqnarray}

We suppose here that the field operators are given in the Schr\"odinger  representation, the spinor components
of the EPF operators being defined in the standard way \cite{Heitler}

\begin{eqnarray}
\label{Eqn19} \hat \psi_{\nu} (\vec{r}) = \sum_{s} \int \frac{d\vec{p}}{(2\pi)^{3/2}} \{a_{\vec{p}s}
u_{\vec{p}s\nu} e^{i\vec{p}\vec{r}} +
b^+_{\vec{p}s} v_{-\vec{p}-s\nu} e^{-i\vec{p}\vec{r}}\};\nonumber\\
\hat \psi^*_{\nu} (\vec{r}) = \sum_{s} \int \frac{d\vec{p}}{(2\pi)^{3/2}} \{a^+_{\vec{p}s} u^*_{\vec{p}s\nu}
e^{-i\vec{p}\vec{r}} + b_{\vec{p}s} v^*_{-\vec{p}-s\nu} e^{i\vec{p}\vec{r}}\}.
\end{eqnarray}

In these formulas $\vec{\alpha}, \beta$ are Dirac matrixes; $u_{\vec{p}s\nu}$ and $v_{\vec{p}s\nu}$ are the
components of the bispinors corresponding to the solutions of  Dirac equations for the free "bare" electron and
positron with the momentum $\vec{p}$ and spin s; $a_{\vec{p}s}(a^+_{\vec{p}s})$ and
$b_{\vec{p}s}(b^+_{\vec{p}s})$ are  the annihilation (creation) operators for the "bare" electrons and positrons
in the corresponding states.  The field operator $\hat{ \vec{A}}(\vec{r})$ and the operator of the photon number
$\hat n_{\vec k \lambda}$ are related to the transversal electromagnetic field and their explicit form will be
written below. It should also be reminded that the parameter  $e_0$ was introduced as the positively defined
value.

We should especially discuss the dependence of the QED hamiltonian on the operator of the scalar field

\begin{equation}
\label{19a} \hat \varphi (\vec{r}) = \sqrt{4\pi}\int d\vec{k} \hat \varphi_{\vec{k}}e^{i\vec{k}\vec{r}} .
\end{equation}

It is known\cite{Heitler}, that these operators could be excluded from the hamiltonian in the Coulomb gauge. For
that purpose one should use the solution of the operator equations of motion for   $\hat \varphi_{\vec{k}}$
supposing that the "bare" electrons are point-like particles and referring to "self-action" is equivalent to the
substitution of the initial mass for the renormalized one. In the result the terms with scalar fields in the
hamiltonian are reduced to the Coulomb interaction between the charged particles. But we can't use this
transformation of the hamiltonian (\ref{Eqn18}) because it is the dynamics of  mass renormalization that we
investigate.

One more problem is connected with a negative sign of the term corresponding to self-energy of the scalar field.
If the non-relativistic problems were considered then the operator of the particle kinetic energy would be
positively defined and the negative operator with the square-law dependence on  $\hat\varphi (\vec{r})$ would
lead to the "fall on the center" \cite{Landau} as the energy minimum would be reached at an infinitely large
field amplitude. However, if the relativistic fermionic field is considered then the operator of the free
particle energy (the first term in formula (\ref{Eqn18})) is not positively definite. Besides, the states of the
system  with the negative energy are filled. Therefore, the stable state of the system corresponds to the energy
extremum (!) (the minimum one for electron and the maximum one for positron excited states). It can be reached
at the finite value of field amplitude (see below). The same reasons permit one to successfully use the states
with indefinite metric \cite{Akhiezer}) in QED although it leads to some difficulties in the non-relativistic
quantum mechanics.

According to our main supposition about the large value of the initial coupling constant  $e_0$ we are to
realize the nonperturbative description of the excited state which is nearest to the vacuum state of the system.
The basic method for an nonperturbative estimation of the energy is the variational approach with some trial
state vector $|\Phi_1 >$ for an approximate description of the one-particle excitation. The qualitative
properties of the self-consistent excitation in the strong coupling limit \cite{Pekar} show that such trial
vector should correspond to the general form of the wave packet formed by the one-particle excitations of the
"bare" EPF. Besides, the effect of polarization and the appearance of the electrostatic field $\varphi
(\vec{r})$ should be taken into account so we consider  $|\Phi_1>$ to be the eigenvector for the operator of the
scalar field. Now, let's introduce the following trial vector depending on the set of variational classical
functions $U_{\vec q s}; V_{\vec q s};\varphi (\vec r)$ for an approximate description of the quasi-particle
excited state of the system:

\begin{eqnarray}
\label{Eqn20} |\Phi_1 > \simeq |\Phi^{(0)}_1(U_{\vec q s}; V_{\vec q s};\varphi (\vec r) )> = \int d \vec{q} \{
U_{\vec{q}s} a^+_{\vec{q} s} + V_{\vec{q}s} b^+_{\vec{q} s} \} | 0;
0;\varphi(\vec r)>; \nonumber\\
\hat \varphi(\vec r)| 0; 0;\varphi(\vec r)> =\varphi(\vec r)| 0; 0;\varphi(\vec r)>; \quad a_{\vec{q} s}| 0;
0;\varphi(\vec r)> = b_{\vec{q} s}| 0; 0;\varphi(\vec r)> = 0.
\end{eqnarray}

The ground state of the system is   $|\Phi_0> = |0;0;0>$, if we use the same notation. It corresponds to the
vacuum of both  EPF and EMF.

Firstly, let's consider the excitation with the zero total momentum. Then the constructed trial vector should
satisfy the normalized conditions resulting from the definition of the total momentum $\vec{P}$ and the observed
charge $e$ of the "physical" particle:

\begin{eqnarray}
\label{Eqn21} <\Phi^{(0)}_1|\hat{ \vec {P}}|\Phi^{(0)}_1> = \sum_{s}\ d \vec{q} \vec{q} [|U_{\vec{q}s}|^2 +
|V_{\vec{q}s}|^2] = \vec{P} = 0;\nonumber\\
\sum_{s}\ d \vec{q} [|U_{qs}|^2 + |V_{qs}|^2] = 1.
\end{eqnarray}

\begin{eqnarray}
\label{Eqn21a} <\Phi^{(0)}_1|\hat Q|\Phi^{(0)}_1> = e_0 \sum_{s}\ d \vec{q} [|V_{qs}|^2 - |U_{qs}|^2] = e.
\end{eqnarray}

The condition (\ref{Eqn21}) restricts the functions $U_{qs}; V_{qs}$ in the sense that they depend on the
modulus of the vector $\vec{q}$ only. It should also be taken into account that the trial vector
$|\Phi^{(0)}_1>$ is not the accurate eigenvector of the exact integrals of motion $\hat Q$ and $\hat{ \vec {P}}$
as it represents the accurate eigenvector of the hamiltonian $|\Phi_1>$ only approximately. Therefore, in the
considered zero approximation the conservation laws for momentum and charge can be satisfied only on the
average, and this leads to the above written normalized conditioned. Generally, the equation  (\ref{Eqn21a})
should not be considered as the additional condition for the variational parameters but as the definition of the
observed charge of the "physical" particle if the initial charge of the "bare" particle is considered as the
given value. Therefore the sign of the observed charge is not fixed a priori,the same was the case for the
qualitative estimation in (\ref{Eqn5}). Calculating the sequential approximations to the exact state vector
$|\Phi_1>$ (see $\S 4$) should restore the accurate integral of motion as well. An analogous problem appeares in
the "polaron" theory when the momentum conservation law was taken into account for the case of the strong
coupling (for example, \cite{Bogol}, \cite{Gross}).

In this respect the variational approach differs essentially from the perturbation theory where the zero
approximation for one-particle state is described by one of the following state vectors:

\begin{eqnarray}
\label{Eqn21b} |\Phi_1> \simeq |\Phi^{(PT)}_1e> = a^+_{\vec{P} s} | 0; 0; 0 >; \quad |\Phi_1> \simeq
|\Phi^{(PT)}_1p> = b^+_{\vec{P} s} | 0; 0; 0 >.
\end{eqnarray}

These vectors don't depend on any parameters and are eigenvectors of the momentum and charge operators. But they
correspond to one-particle excitations determined by the charge $e_0$ of the "bare" electron and the field
$\varphi (\vec r ) = 0$. As supposed the introduction of the variational parameters into the wave function of
the zero approximation enabled us to take into account the vacuum polarization, but the variational approach in
this form allows to conserve the exact integral of motion only on the average. This problem will be considered
in more detail below (Section 4).

So, the following variational estimation for the energy $E_1 (\vec{P}=0)$ of the state corresponding to the
"physical" quasi-particle excitation of the whole system is considered in the strong coupling zero
approximation:

\begin{eqnarray}
\label{Eqn22} E_1 (0) \simeq  E^{(0)}_1 [U_{qs}; V_{qs};\varphi(\vec{r})] = <\Phi^{(0)}_1|\hat H |\Phi^{(0)}_1>,
\end{eqnarray}

where the average is calculated with the full hamiltonian (\ref{Eqn18}) and the functions $U_{qs};
V_{qs};\varphi(\vec{r})$ are to be found as the solutions of variational equations

\begin{eqnarray}
\label{Eqn23} \frac{\partial E^{(0)}_1 (U_{qs}; V_{qs};\varphi(\vec{r}))}{\partial U_{qs}} = \frac{\partial
E^{(0)}_1 }{\partial V_{qs}} =  \frac{\partial E^{(0)}_1 }{\partial \varphi(\vec{r})} = 0
\end{eqnarray}

under some additional conditions (\ref{Eqn21} - \ref{Eqn21a}).

It is quite natural, that the ground state energy is calculated  in the framework of the considered
approximation as follows

\begin{eqnarray}
\label{Eqn24} E_0 \simeq  E^{(0)}_0 = <\Phi_0|\hat H |\Phi_0>.
\end{eqnarray}

Then the energy of the quasi-particle representing the "physical" electron (positron) is calculated by the
formula

\begin{eqnarray}
\label{Eqn24a} E(\vec P= 0) = E^{(0)}_1 -  E^{(0)}_0.
\end{eqnarray}

Further discussion is needed in connection with the application of the variational principle
(\ref{Eqn22}-\ref{Eqn23}) for estimating the energy of the excited state, because usually the variational
principle is used for estimating the ground state energy. As far as we knew Caswell \cite{Caswell} was the first
to apply such method for nonperturbative calculation of the excited states of the anharmonic oscillator with an
arbitrary value of the anharmonicity . He  called this approach as "principle of the minimal sensitivity". It
was shown in our paper  \cite{OM82}, that the application of the variational principle to the excited states is
actually the consequence of the fact that the set of eigenvalues for the full hamiltonian does't depend on the
choice of the representation for eigenfunctions. As the result, the operator method (OM) for solving
Schr\"odinger equation was developed as the regular procedure for calculating further corrections to the
zero-order approximation. Later OM was applied to a number of real physical systems and proved to be very
effective when calculating the energy spectrum in the wide range of the hamiltonian parameters and quantum
numbers (\cite{OM95}, \cite{OM96}, \cite{Acta} and the cited references). These results can serve as the
background for the  application of the variational estimation to the excited state energy in QED. The procedure
for calculating  the successive approximations is considered in Section 5.

The average value in Eq. (\ref{Eqn22}) is calculated neglecting the classical components of the vector field
(quantum fluctuations should be taken into account in the further approximations)

\begin{eqnarray}
\label{Eqn25} <\Phi^{(0)}_1 | \hat \psi^* (\vec{r}) [ \vec \alpha \hat{ \vec{ A} } (\vec{r})] \hat \psi
(\vec{r})|\Phi^{(0)}_1> = 0.
\end{eqnarray}

It should be noted that the possibility of constructing self-consistently the renormalized QED at the non-zero
vacuum value of the scalar field operator was considered before  \cite{Fradkin} but the solution itself of the
corresponding equations was not  discussed.

It is also essential that the normal ordering of the operators  \cite{Bogoliubov} wasn't introduced in the
hamiltonian (\ref{Eqn18}). Therefore, its average value includes the infinite values corresponding to the vacuum
energies of EPF for both  the ground and the excited states. However, if the normalized conditions (\ref{Eqn21})
are fulfilled these values are compensated in the expression (\ref{Eqn24a}) for the quasi-particle energy. Then
the functional for determining the zero approximation for the energy of the EPF one-particle excitation is
defined as follows:

\begin{eqnarray}
\label{Eqn25a} E(\vec P=0) = \int d \vec r \int \frac{d \vec q}{(2\pi)^{3/2}} \int \frac{d \vec
{q}'}{(2\pi)^{3/2}} \sum_{s,s'} \sum_{\mu,\nu} \{ U^*_{q's'} u^*_{\vec{q}' s' \mu} [(\vec \alpha \vec q + \beta
m_0)_{\mu \nu} + \nonumber\\
e_0 \varphi (\vec r) \delta_{\mu \nu}]
U_{q s} u_{\vec{q} s \nu} - \nonumber\\
V_{q's'} v^*_{\vec{q}' s' \mu} [(\vec \alpha \vec q + \beta m_0)_{\mu \nu} + e_0 \varphi (\vec r) \delta_{\mu
\nu}] V^*_{qs} v_{\vec{q} s \nu}\} e^{i (\vec q - \vec{q}') \vec r} - \nonumber\\
\frac{1}{2} \int d \vec r
[\vec{\nabla} \varphi (\vec r)]^2.
\end{eqnarray}

In order to vary the introduced functional let us define the spinor wave functions (not operators) which
describe the coordinate representation for the electron and positron wave packets in the state vector  $
|\Phi^{(0)}_1>$:

\begin{eqnarray}
\label{Eqn26} \Psi_{\nu} (\vec r) = \int \frac{d \vec q}{(2\pi)^{3/2}} \sum_{s} U_{q s} u_{\vec{q} s \nu}
e^{i \vec q \vec r}; \nonumber\\
\Psi^c_{\nu} (\vec r) = \int \frac{d \vec q}{(2\pi)^{3/2}} \sum_{s} V^*_{q s} v_{\vec{q} s \nu} e^{i \vec q \vec
r}.
\end{eqnarray}

In particular, if the trial state vector is chosen in one of the forms (\ref{Eqn21b}) of the standard
perturbation theory, the wave functions  (\ref{Eqn26}) coincide with the plane wave solutions of the free Dirac
equation. For a general case the variation of the functional (\ref{Eqn25a}) by the scalar field leads to

\begin{eqnarray}
\label{Eqn27} E(0) = \int d \vec r \{ \Psi^+ (\vec r) [(-i\vec \alpha \vec \nabla + \beta m_0) +
\frac{1}{2}e_0 \varphi (\vec r) ] \Psi (\vec r) - \nonumber\\
\Psi^{+c} (\vec r) [(-i\vec \alpha \vec \nabla + \beta m_0) +
\frac{1}{2}e_0 \varphi (\vec r) ] \Psi^c (\vec r); \nonumber\\
\int {d \vec{r}} [\Psi^{+} (\vec r) \Psi (\vec r) + \Psi^{+c} (\vec r') \Psi^{c} (\vec r')] = 1;
\end{eqnarray}

\begin{eqnarray}
\label{Eqn27b} \varphi (\vec r) = \frac{e_0}{4 \pi} \int \frac {d \vec{r}'} {|\vec r - \vec{r}'|} [\Psi^{+}
(\vec r') \Psi (\vec r') - \Psi^{+c} (\vec r') \Psi^{c} (\vec r')]; \nonumber\\
\int {d \vec{r}} [\Psi^{+} (\vec
r) \Psi (\vec r) + \Psi^{+c} (\vec r') \Psi^{c} (\vec r')] = 1.
\end{eqnarray}

The principal condition for the existence of the considered nonperturbative excitation in QED is defined by the
minimum of the functional (\ref{Eqn27}) corresponding to a non-zero classical field. The structure of this
functional shows that such solutions of the variational equations could appear only if the trial state vector
included the superposition of the electron and positron wave packets simultaneously.

The functional (\ref{Eqn27}) is the analogue of the quasi-classical expression (\ref{Eqn5}). Equation
(\ref{Eqn27}) and the Fourier representation  (\ref{Eqn20}) for the trial vector clearly indicate that the
supposition concerning the localization of the functions $\Psi (\vec r)$ near some point doesn't contradict the
translational symmetry of the system because this point itself can be situated at any point of the full space
with equal probability. The general analysis of the correlation between the local violation of the symmetry and
the conservation of accurate integral of motion for the arbitrary quantum system was first discussed in detail
by Bogoluibov in his widely known paper "On quasi-averages" \cite{quasi}. Analogous analysis in the considered
problem we'll discuss in Section 4.

Varying the functional (\ref{Eqn27}) by the wave functions $\Psi (\vec r)$ è $\Psi^c (\vec r)$ taking into
account their normalization leads to the Dirac equations in their usual form describing the electron (positron)
motion in the field of  potential $\varphi (\vec r)$:

\begin{eqnarray}
\label{Eqn27a} \{(-i\vec \alpha \vec \nabla + \beta m_0) +
e_0 \varphi (\vec r) - E(0)\} \Psi (\vec r) = 0;\nonumber\\
\{(-i\vec \alpha \vec \nabla + \beta m_0) + e_0 \varphi (\vec r) + E(0)\} \Psi^{c} (\vec r) = 0.
\end{eqnarray}

We should discuss the procedure of separatiing variables in more detail, because of the non-linearity of the
obtained system of  equations for the wave functions and the self-consistent potential. Since the considered
physical system has no preferred vectors if  $\vec P = 0$, it is natural to regard the self-consistent potential
as spherically symmetrical. Then the variable separation for the Dirac equation is realized on the basis of the
well known spherical spinors \cite{Akhiezer}:

\begin{eqnarray}
\label{Eqn28}
\Psi_{jlM} = \left( \begin{array}{c}
g(r) \Omega_{jlM}\\
i f(r) \Omega_{jl'M}
\end{array} \right).
\end{eqnarray}

Here $\Omega_{jlM}$ are the spherical spinors \cite{Akhiezer} describing the spin and angular dependence of the
one-particle excitation wave functions; $j, M$ are the total excitation momentum and its projection; the orbital
momentum eigenvalues are connected by the correlation $ l + l' = 2j$. It is natural to consider the state with
the minimal energy as the most symmetrical one, corresponding to the values  $j=1/2; M = \pm 1/2; l = 0; l' =
1$. This choice corresponds to the condition that in the non-relativistic limit the "large" component of the
spinor $\Psi$ $\sim g$ corresponds to the EPF electronic zone . In this case the unknown radial functions $f, g$
satisfy the following system of the equations:

\begin{eqnarray}
\label{Eqn29} \frac{d (rg)}{dr} - \frac{1}{r}(rg) - (E + m_0 - e_0 \varphi(r)) (rf) = 0;
\nonumber\\
\frac{d (rf)}{dr} + \frac{1}{r}(rf) + (E - m_0 - e_0 \varphi(r)) (rg) = 0.
\end{eqnarray}

The states with various projections of the total momentum should be equally populated in order to be consistent
with the supposition of the potential spherical symmetry with the equation (\ref{Eqn27b}). So, the total wave
function of the "electronic" component of the EPF quasi-particle excitation is chosen in the following form:

\begin{eqnarray}
\label{Eqn28a} \Psi = \frac{1}{\sqrt{2}} [\Psi_{1/2,0,1/2} + \Psi_{1/2,0,-1/2}]=
\left(
\begin{array}{c}
g(r) \chi^+_0\\
i f(r) \chi^+_1
\end{array} \right);
\nonumber\\
\chi^+_l =\frac{1}{\sqrt{2}} [\Omega_{1/2,l,1/2} + \Omega_{1/2,l,-1/2}]; \quad l = 0;1.
\end{eqnarray}

In its turn, the wave function  $\Psi^c$ is defined on the basis of the following bispinor:

\begin{eqnarray}
\label{Eqn30} \Psi^c_{jlM} = \left( \begin{array}{c}
- i f_1(r) \Omega_{jlM}\\
g_1(r) \Omega_{jl'M}
\end{array} \right).
\end{eqnarray}

The radial wave functions  $f_1, g_1$ in this case satisfy the following system of equations

\begin{eqnarray}
\label{Eqn31} \frac{d (rg_1)}{dr} + \frac{1}{r}(rg_1) - (E + m_0 + e_0 \varphi(r)) (rf_1) = 0;
\nonumber\\
\frac{d (rf_1)}{dr} - \frac{1}{r}(rf_1) + (E - m_0 + e_0 \varphi(r)) (r g_1) = 0.
\end{eqnarray}

These equations correspond to  the positronic branch of the EPF spectrum with the "large" component  $\sim g_1$
in the non-relativistic limit.

It is important to note that the functions  $\Psi$ and $\Psi^c$ satisfy the equations (\ref{Eqn27}) with the
same energy E(0). It imposes an additional condition of the orthogonality to them:

\begin{eqnarray}
\label{Eqn31a} <\Psi^c|\Psi>  = 0.
\end{eqnarray}

Taking into account this condition and also the requirement that the states with  different values of M should
be equally populated one finds the "positronic" wave function

\begin{eqnarray}
\label{Eqn31b} \Psi^c = \frac{1}{\sqrt{2}} [\Psi^c_{1/2,0,1/2} - \Psi^c_{1/2,0,-1/2}]= \left(
\begin{array}{c}
-i f_1(r) \chi^-_0\\
g_1(r) \chi^-_1
\end{array} \right );
\nonumber\\
\chi^-_l =\frac{1}{\sqrt{2}} [\Omega_{1/2,l,1/2} - \Omega_{1/2,l,-1/2}]; \quad l = 0;1.
\end{eqnarray}

The equation for the self-consistent potential follows from the definition of $\varphi (r)$ in  formula
 (\ref{Eqn27a})taking into account the normalization of the spherical spinors  \cite{Akhiezer}:

\begin{eqnarray}
\label{Eqn32} \frac{d^2 \varphi}{d r^2} + \frac{2}{r}\frac{d \varphi}{d r} = - \frac{e_0}{4\pi}[f^2 + g^2 -
f_1^2 - g_1^2].
\end{eqnarray}

The boundary condition for the potential is equivalent to the normalization condition (\ref{Eqn21a}) and defines
the charge of the "physical" electron (positron) e

\begin{eqnarray}
\label{Eqn33}
\varphi (r)|_{r \rightarrow \infty} = \frac{e}{4 \pi r} = \nonumber\\
\frac{e_0}{4\pi r}\int_{0}^{\infty} r^2_1 dr_1[f^2(r_1) + g^2(r_1) - f_1^2(r_1) - g_1^2(r_1)].
\end{eqnarray}

It is important to stress that the form of the functions given above is defined practically uniquely by the
imposed conditions. At the same time the obtained equations are consistent with the symmetries defined by the
physical properties of the system. The first one is quite evident and relates to the fact that the excitation
energy doesn't depend on the choice of the quantization axis of the total angular momentum.

Moreover, these equations satisfy the condition of the charge symmetry \cite{Akhiezer}. Indeed, one can check
directly that one more pair of bispinors leads to the equations completely coinciding with
(\ref{Eqn29}),(\ref{Eqn31})

\begin{eqnarray}
\label{Eqn33a}\tilde{ \Psi}_{jlM} = \left( \begin{array}{c}
i g_1(r) \Omega_{jl'M}\\
- f_1(r) \Omega_{jlM}
\end{array} \right);
\end{eqnarray}

\begin{eqnarray}
\label{Eqn33b}\tilde{ \Psi}^c_{jlM}= \left( \begin{array}{c}
- f(r) \Omega_{jl'M}\\
- i g(r) \Omega_{jlM}
\end{array} \right).
\end{eqnarray}

It means that these bispinors allows one to find another pair of the wave functions which are orthogonal to each
other and to the functions  (\ref{Eqn28a}),(\ref{Eqn31b}) but including the same set of the radial functions

\begin{eqnarray}
\label{Eqn33ñ}
\tilde{\Psi} = \left(
\begin{array}{c}
i g_1(r) \chi^-_1\\
-f_1(r) \chi^-_0
\end{array} \right);\quad
\tilde{\Psi}^c =\left(
\begin{array}{c}
- f(r) \chi^+_1\\
-i g(r) \chi^+_0
\end{array} \right).
\end{eqnarray}

These functions differ from the set (\ref{Eqn28a}),(\ref{Eqn31b}) in that they have another sign of the observed
charge of the quasi-particle due to the boundary condition  (\ref{Eqn33}).

Let us now proceed to the solution of the variational equations. It follows from the  qualitative analysis that
the important property of the trial state vector freedom is the relative contribution of the electronic and
positronic components of the wave function. Therefore let us introduce the variational parameter C in the
following way:

\begin{eqnarray}
\label{Eqn34}
\int_{0}^{\infty} r^2 dr[f^2(r) + g^2(r)] = \frac{1}{1 + C}; \nonumber\\
\int_{0}^{\infty} r^2 dr[f_1^2(r) + g_1^2(r)] = \frac{C}{1 + C}.
\end{eqnarray}

The dimensionless variables and new functions can be introduced

\begin{eqnarray}
\label{Eqn35} x = r m_0; \quad E = \epsilon m_0; \quad  e_0 \varphi(r) = m_0\phi(x);
\quad \frac{e^2_0}{4 \pi} = \alpha_0;\nonumber\\
u(x) \sqrt{m_0} = r g(r);\quad v(x)\sqrt{m_0} = r f(r);\quad u_1(x)\sqrt{m_0} = r g_1(r);\quad v_1(x)\sqrt{m_0}
= r f_1(r).
\end{eqnarray}

As the result the system of equations for describing the radial wave functions of the EPF one-particle
excitation and the self-consistent potential of the vacuum polarization can be obtained.  These equations can be
clearly interpreted and the equations  (\ref{Eqn9} - \ref{Eqn11}) serve as their quasi-classical analogue.

\begin{eqnarray}
\label{Eqn36} \frac{d u}{dx} - \frac{1}{x}u - (\epsilon + 1 - \phi(x)) v  = 0;
\nonumber\\
\frac{d v}{dx} + \frac{1}{x}v + (\epsilon - 1 - \phi(x)) u = 0;
\nonumber\\
\frac{d u_1}{dx} + \frac{1}{x}u_1 - (\epsilon + 1 + \phi(x)) v_1 = 0;
\nonumber\\
\frac{d v_1}{dx} - \frac{1}{x}v_1 + (\epsilon - 1 + \phi(x)) u_1 = 0;
\nonumber\\
\phi(x) = \alpha_0 [ \int_{x}^{\infty} dy \frac{\rho(y)}{y} +
\frac{1}{x} \int_{0}^{x} dy \rho(y)];\nonumber\\
\rho(x) = [u^2(x) + v^2(x) - u^2_1(x) - v^2_1(x)].
\end{eqnarray}

These equations should be complemented by the conditions which correspond to the relations between the observed
charge and mass, by means of their initial values and eigenvalues  $\epsilon, C$ of the equations (\ref{Eqn36}):

\begin{eqnarray}
\label{Eqn37}
E(0) = \epsilon m_0 = \pm m;\nonumber\\
e_0 \frac{(1 - C)}{(1 + C)}  = \pm |e|.
\end{eqnarray}

The sign under the renormalization conditions (\ref{Eqn37}) will depend on the correlation between the value
$e_0$, calculated on the basis of the equations (\ref{Eqn36}) and the critical parameter $e^*_0$ as it was
discussed in Section 2.

The mathematical structure of the non-linear equations (\ref{Eqn36}) is in many respects similar to that of the
equations determining the ground state of the system in the "polaron" problem \cite{Pekar}, but the number of
the equations in our case is larger. Generally speaking, we can find the direct numerical solution of these
equations (\cite{Krylov}). But in the present paper we are restricted by an approximate variational solution
which is usually close to the exact solution in the similar problem  \cite{Pekar}. In order to find this
solution let us use the functional which is an exact consequence of the functional (\ref{Eqn27}) for E(0):

\begin{eqnarray}
\label{Eqn38} I = \int_{0}^{\infty} dx \{ (u' v - v' u) - 2 \frac{u v}{x} -
(\epsilon + 1 - \phi) v^2 - (\epsilon - 1 - \phi) u^2 +\nonumber\\
(u_1' v_1 - v_1' u_1) + 2 \frac{u_1 v_1}{x} - (\epsilon + 1 + \phi) v_1^2 - (\epsilon - 1 + \phi) u_1^2\} = 0.
\end{eqnarray}

The potential  $\phi(x)$ in the functional (\ref{Eqn36}) is supposed to be calculated as the exact solution of
the Poisson equation, as it was  made by Pekar  \cite{Pekar1}. The choice of the form for trial functions is
defined by the conditions of the analytical behavior for the solutions in the limit $x \rightarrow 0$ and their
asymptotic for $x \rightarrow \infty$. The simplest set of the functions satisfying these conditions and
normalized in accordance with  (\ref{Eqn34}) is as follows:

\begin{eqnarray}
\label{Eqn39}
u(x) = 2A \sqrt{\frac{b^3}{7(1+ C )}} (1 + b x) x e^{-bx};\nonumber\\
v(x) = 2b x^2 \sqrt{\frac{(1 - A^2)b^3}{3(1+ C )}} e^{-bx};\nonumber\\
u_1(x) = 2b x^2 \sqrt{\frac{(C - A_1^2)b^3}{3(1 + C )}} e^{-bx};\nonumber\\
v_1(x) = 2A_1 \sqrt{\frac{b^3}{7(1 + C )}} (1 + b x) x e^{-bx}.
\end{eqnarray}

Unlike the quasi-classical estimation  (\ref{Eqn6}) this set of  functions depends on 4 variational parameters
($A, A_1, C, b$) with quite evident physical interpretation. The self-consistent potential is calculated by
means of the last equation in (\ref{Eqn36}):

\begin{eqnarray}
\label{Eqn40} \phi(x) = \frac{\alpha_0}{x(1+Ñ)}\{(1 - C) - \frac{1}{42}e^{-2bx}[ 42 (1 - C) + 3 b x (21 - 2 A^2
+ 2 A_1^2 - 21 C) + \nonumber\\6 b^2 x^2 (7 - 2 A^2 + 2 A_1^2 - 7 C) + 2 b^3 x^3 (7 - 4 A^2 + 4 A_1^2 - 7 C)]\}.
\end{eqnarray}

As the result the functional (\ref{Eqn38}) transforms as follows

\begin{eqnarray}
\label{Eqn41} I = (C-1) - \epsilon (C + 1) + 2 b \sqrt{\frac{3}{7}} (A_1 \sqrt{C - A^2_1} -
A \sqrt{1 - A^2}) + 2 (A^2 - A_1^2) + \nonumber\\
b \frac{\alpha_0}{C+1} [ \frac{3}{56}(A^2 - A_1^2)(1 - C) + \frac{93}{256}(1 - C)^2 + \frac{9}{784}(A^2 -
A^2_1)^2 ] = 0.
\end{eqnarray}

In a general case the variation of the dimensionless energy $\epsilon$ in equation (\ref{Eqn41}) by 4 parameters
leads to rather complicated equations. Their numerical solution depends on the value $\alpha_0$ which, in its
turn,  should be calculated from the renormalization condition (\ref{Eqn37}). However, the observed charge of
the "physical" electron is defined by the parameter $\alpha_0 \gg 1$ (see the final result and the
quasi-classical estimation). In this limit the unknown solution are simplified and can be represented in the
analytical form

\begin{eqnarray}
\label{Eqn42} A^2 - A_1^2 \simeq \frac{2t}{\alpha_0}; \qquad (1 - C) \simeq \frac{2z}{\alpha_0};
\nonumber\\
\epsilon = \frac{q(t,z,A,b)}{\alpha_0},
\end{eqnarray}

\noindent where the introduced  parameters are defined in terms of the conditions of numerical functional
minimum $q(t,z,A,b)$ which doesn't depend on the coupling constant $\alpha_0$ for this approximation:

\begin{eqnarray}
\label{Eqn43} q(t,z,A,b) = 2 t - z - b\sqrt{\frac{3}{7(1 - A^2)}}[\frac{t(1 - 2A^2)}{A} + A z] +
\nonumber\\
b (\frac{9 t^2}{784} + \frac{3tz}{56} + \frac{93z^2}{256})
\end{eqnarray}

When 3 parameters are excluded from the corresponding variational equations the algebraic equation for the last
one can be obtained

\begin{eqnarray}
\label{Eqn44} 1519 - 12376 A^2 + 11560 A^4 + 1088 A^8 = 0.
\end{eqnarray}

A numerical solution of this equation allows to calculate the other parameters and the value of the functional
$q = q_0$ at the point of minimum

\begin{eqnarray}
\label{Eqn45}
A = - 0.920843; \quad b = 1.48024; \quad z = -1.12751;\nonumber\\
t = - 0.956075; \quad q_0 = - 0.784642.
\end{eqnarray}

It can be noted that the choice of a negative sign for A value follows from the condition that the parameter b
should be positive.

The signs of the parameters  $z, q_0$ show that properties of the quasi-particle resulting from the one-electron
excitation of EPF correspond to the "physical" positron as it was obtained for the quasi-classical analysis. But
the numerical values of the initial QED parameters expressed through the observed values by means of Eqs.
(\ref{Eqn30}) slightly differ from (\ref{Eqn17}):

\begin{eqnarray}
\label{Eqn46} \alpha_0 =\frac{z^2}{\alpha}\simeq 174; \quad \alpha = \frac{e^2}{4\pi}
\simeq \frac{1}{137}; \nonumber\\
m_0 = \frac{\alpha_0}{|q_0|} m \simeq 222 m \simeq 113 (MeV);\nonumber\\
r^0_e = \frac{1}{m_0 b} = \frac{|q_0|}{ b z^2} \frac{\alpha}{m} \simeq 0.416 r_e.
\end{eqnarray}

The last equation shows that the "electromagnetic radius"  $r^0_e$ of the quasi-particle is of the same order of
magnitude as the parameter for the "classical electron radius"  $r_e$ in the renormalized QED. Besides, it is
worth mentioning that the mass of "bare" electron seems to be close to the observed mass of muon (106 MeV).

\section{EPF quasi-particle excitation with arbitrary momentum}

In the previous section we have considered  the possibility for the resting quasi-particle with a non-trivial
self-consistent charge distribution, the finite energy $E(0)$ and a zero total momentum $\vec P = 0$ to exist in
the framework of nonperturbative QED.

The obtained solution allows one to imagine the internal structure of the resting "physical" electron (positron)
as the strongly coupled state of charge distributions of the opposite sign, the large values of integral charges
of these distributions compensating each other almost completely and their heavy masses are being "absorbed" by
the binding energy.

Actually the energy $\pm E(0)$ defines the boundaries of the renormalized electron and positron zones resulting
from the strong polarization of EPF when the excitation appears  (Figure 1). But this excitation could be really
interpreted as the "physical" electron (positron) if the sequence of the levels in every zone determined by the
vector $\vec P \not= 0 $ (Fig.1) is described by the relativistic energy spectrum of real particles, that is

\begin{eqnarray}
\label{Eqn47} E(\vec P) = \sqrt{ P^2 + E^2 (0) } = \sqrt{ P^2 + m^2 }.
\end{eqnarray}

Only in this case the energy E(0) can really be used for  calculating the mass $m_0$ of the "bare" electron.

It is worth saying, that the problem of investigating the dynamics of the self-localized excitation should be
solved for any system with the strong interaction between quantum fields in order to calculate its effective
mass. For example, a similar problem for Pekar "polaron"   \cite{Pekar} in the ionic crystal was considered in
the papers  \cite{PV},\cite{Bogol}, \cite{Tyablikov}, \cite{Feynmanpol}, \cite{Gross} and in many more recent
works. It is essential that because of the non-linear coupling between the particle and a self-consistent field
the energy dispersion $E(\vec P)$ for the quasi-particle proved to be very complicated. As the result, its
dynamics in the crystal is similar to the motion of the point  "physical" particle only at rather small full
momentum.

However, in the case of QED the problem is formulated fundamentally in a different way. At present there is no
doubt  that the dynamics of the "physical" electron is described by the formula (\ref{Eqn47}) for any (!) values
of the momentum $\vec P$. It means that the considered nonperturbative approach for describing the internal
structure of the "physical" electron should lead to the energy dispersion law (\ref{Eqn47}) for the entire range
of the momentum $\vec P$. The analysis of this problem can be considered as an important test for finding out
whether our method suits the experimental data.

In order to describe the quasi-particle excitation with the non-zero momentum in the framework of the
variational principle, let us introduce a more general trial state vector than the one in the formula
(\ref{Eqn20})

\begin{eqnarray}
\label{Eqn47a} |\Phi_1(\vec P) > \simeq |\Phi^{(0)}_1(U_{\vec q,\vec P,s}; V_{\vec q,\vec P,s};\varphi(\vec r,
\vec P); \vec{P} )> = \nonumber\\
\int d \vec{q} \{ U_{|\vec{q} - \vec{P}|s} a^+_{\vec{q} s} + V_{|\vec{q} -
\vec{P}|s} b^+_{\vec{q} s} \} |0; 0;\varphi(\vec r, \vec P); \vec{P} >.
\end{eqnarray}

We should stress once more that the wave packet   (\ref{Eqn47a}) doesn't distinguish any point in the space when
the energy of the stationary state is estimated by the variational functional. The matter is that the shift of
the coordinate in the self-consistent potential   $\varphi(\vec r, \vec P)$ is reduced to the change of the
variable when integrating in the full QED hamiltonian. The constructed state vector  differs from the
one-particle wave packet  (\ref{Eqn20})in that the momenta of the EPF excitations forming the internal structure
of the "physical" electron are counted off from the vector  $\vec P$. The variational parameters  $U_{\vec
q,\vec P,s};V_{\vec q,\vec P,s}$ depend only on the modulus of the vectors difference because of the
normalization condition and the conservation on the average of the integral of motion

\begin{eqnarray}
\label{Eqn47b} <\Phi^{(0)}_1|\hat{ \vec {P}}|\Phi^{(0)}_1> = \sum_{s}\ d \vec{q} \vec{q} [|U_{|\vec{q} -
\vec{P}|s}|^2 +
|V_{|\vec{q} - \vec{P}|s}|^2] = \vec{P};\nonumber\\
\sum_{s}\ d \vec{q} [|U_{qs}|^2 + |V_{qs}|^2] = 1.
\end{eqnarray}

\begin{eqnarray}
\label{Eqn47c} <\Phi^{(0)}_1|\hat Q|\Phi^{(0)}_1> = e_0 \sum_{s}\ d \vec{q} [|V_{qs}|^2 - |U_{qs}|^2] = e.
\end{eqnarray}

Then the analogue of the functional (\ref{Eqn25a}) determining the energy  $E(\vec P) = E^{(0)}_1 - E^{(0)}_0$
of the quasi-particle excitation with an arbitrary momentum is defined as follows

\begin{eqnarray}
\label{Eqn47d} E(\vec P) = \int d \vec r \int \frac{d \vec q}{(2\pi)^{3/2}} \int \frac{d \vec
{q}'}{(2\pi)^{3/2}} \sum_{s,s'} \sum_{\mu,\nu} \{ U^*_{|\vec {q}' - \vec P| s'} u^*_{\vec{q}' s' \mu} [(\vec
\alpha \vec q + \beta m_0)_{\mu \nu} + \nonumber\\
e_0 \varphi (\vec r,\vec P) \delta_{\mu \nu}]
U_{|\vec {q} - \vec P| s} u_{\vec{q} s \nu} - \nonumber\\
V_{|\vec {q}' + \vec P| s'} v^*_{\vec{q}' s' \mu} [(\vec \alpha \vec q + \beta m_0)_{\mu \nu} + e_0 \varphi
(\vec r,\vec P) \delta_{\mu \nu}] V^*_{|\vec {q} + \vec P| s} v_{\vec{q} s \nu}\} e^{i (\vec q - \vec{q}') \vec
r} - \nonumber\\\frac{1}{2} \int d \vec r [\vec{\nabla} \varphi (\vec r,\vec P)]^2.
\end{eqnarray}

Let us introduce  new variational functions in the coordinate representation

\begin{eqnarray}
\label{Eqn48} \Psi_{\nu} (\vec r, \vec P) = \int \frac{d \vec q}{(2\pi)^{3/2}} \sum_{s} U_{|\vec q - \vec P|, s}
u_{\vec{q} s \nu}
e^{i \vec q \vec r}; \nonumber\\
\Psi^c_{\nu} (\vec r) = \int \frac{d \vec q}{(2\pi)^{3/2}} \sum_{s} V^*_{|\vec q + \vec P|, s} v_{\vec{q} s \nu}
e^{i \vec q \vec r}.
\end{eqnarray}

In order to clarify the physical meaning of these functions let's note that if the internal degrees of freedom
described by the bispinors  $u_{\vec{q} s \nu}; v_{\vec{q} s \nu}$ were not taken into account, the wave
function $\Psi (\vec r, \vec P)$ could be represented as

$$
\Psi (\vec r, \vec P) =  e^{i\vec r \vec P}\Psi [(\vec r ), \vec P = 0)].
$$

However, in the problem under consideration we can't use this representation directly because of the mixing up
of  translational and spin variables.

So, the variation of the functional  (\ref{Eqn47d}) by the classical field leads to

\begin{eqnarray}
\label{Eqn49} E(\vec P) = \int d \vec r \{ \Psi^+ (\vec r, \vec P) [(\vec \alpha \vec P -i\vec \alpha \vec
\nabla + \beta m_0) + e_0\frac{1}{2} \varphi (\vec r, \vec P) ] \Psi (\vec r, \vec P) -
\nonumber\\
\Psi^{+c} (\vec r, \vec P) [(- \vec \alpha \vec P -i\vec \alpha \vec \nabla + \beta m_0) +
e_0\frac{1}{2} \varphi (\vec r, \vec P) ] \Psi^c (\vec r, \vec P); \nonumber\\
\varphi (\vec r, \vec P) = e_0 \int \frac {d \vec{r}'} {|\vec r - \vec{r}'|} [\Psi^{+} (\vec r', \vec P) \Psi
(\vec r', \vec P) - \Psi^{+c} (\vec r', \vec P) \Psi^{c} (\vec r', \vec P)];
\nonumber\\
\int {d \vec{r}} [\Psi^{+} (\vec r, \vec P) \Psi (\vec r, \vec P) + \Psi^{+c} (\vec r', \vec P) \Psi^{c} (\vec
r', \vec P)] = 1.
\end{eqnarray}

As it was mentioned above, there is quite a close analogy between the nonperturbative description of QED and the
theory of strong coupling in the "polaron" problem. Therefore it is of interest to compare the obtained
functional (\ref{Eqn49})with the results of various methods of taking into account the translational motion in
the "polaron" problem. The simplest one was used by Landau and Pekar \cite{PV} who introduced the Lagrange
multipliers in the form

\begin{eqnarray}
\label{Eqn49a} J(\vec P) = J( \vec P =0) + (\vec P \vec V),
\end{eqnarray}

with the functional $J( \vec P =0)$ referring to the resting "polaron" and the Lagrange multiplier  $V_i$ having
the meaning of the components of the quasi-particle average velocity.

We can see that the obtained functional(\ref{Eqn49}) has the same form as  (\ref{Eqn49a})if the relativistic
velocity of the excitation is determined by the formula

\begin{eqnarray}
\label{Eqn49b} \vec V = \int d \vec r \{\Psi^+ (\vec r, \vec P) (\vec \alpha)  \Psi (\vec r, \vec P) + \Psi^{+c}
(\vec r, \vec P) (\vec \alpha)  \Psi^c (\vec r, \vec P)\} .
\end{eqnarray}

It corresponds to the well known interpretation of  Dirac matrixes\cite{Dirac}.

Varying this functional by taking into account the normalized conditions leads to the following equations for
the wave functions

\begin{eqnarray}
\label{Eqn50} \{(\vec \alpha \vec P -i\vec \alpha \vec \nabla + \beta m_0) + e_0 \varphi (\vec r, \vec P) -
E(\vec P)\} \Psi (\vec r, \vec P) = 0;
\nonumber\\
\{(\vec \alpha \vec P + i\vec \alpha \vec \nabla - \beta m_0) - e_0 \varphi (\vec r, \vec P) - E(\vec P) \} \Psi^c
(\vec r, \vec P)= 0.
\end{eqnarray}

These equations show that  distinct from the "polaron" problem \cite{Bogol} the translational motion of the
quasi-particle with the momentum $\vec P$ in our case is related to its internal movement described by the
coordinate $\vec r$ only by means of spinor variables only. The physical reason for such separation of variables
is explained by the fact that in QED the self-localized state is formed by the scalar field, and its interaction
with the particle doesn't include the momentum exchange. In order to find the analytical energy spectrum $E(\vec
P)$ the system of non-linear equations(\ref{Eqn50} should be diagonalized relative to the spinor variables. The
possibility of such diagonalization seems to be a non-trivial requirement for the nonperturbative QED under
consideration.

The solution of the equations  (\ref{Eqn50}) can be found on the basis of the states for which the dependence on
the vector $\vec q$ in the wave packet amplitudes  $U_{\vec q, s}, V_{\vec q, s}$ remains the same as it was in
the motionless "physical" electron. However, the relation between the spinor components of these functions can
be changed. But as the self-consistent scalar potential includes the summation over all spinor components it can
be supposed that the potential doesn't depend on the momentum for the class of states in question:

\begin{eqnarray}
\label{Eqn51} \varphi (\vec r, \vec P) = \varphi (r) |_{\vec P = 0}.
\end{eqnarray}

In the coordinate representation the transformation of the spinor components of the wave functions satisfying
the Equations (\ref{Eqn50}) takes place because of the dependence on the momentum. It is possible that there are
some general arguments based on the hamiltonian symmetry that make easier the diagonalization of Equations
(\ref{Eqn50}). However, we have constructed the solution by exhausting  various linear combinations of the wave
functions found in Section 3 for a resting electron. These functions correspond to the degenerated states in the
case of $\vec P = 0$ but are mixed for a moving electron. It was found that there is the only normalized linear
combination satisfying to all the necessary conditions of self-consistency :

\begin{eqnarray}
\label{Eqn52} \Psi (\vec r, \vec P)  = L (\vec P) \Psi (\vec r) + K (\vec P) \tilde{\Psi^c} (\vec r) ;
\nonumber\\
\Psi^c(\vec r, \vec P)  = L_1(\vec P) \Psi^c (\vec r) +
K_1(\vec P) \tilde{\Psi}(\vec r); \nonumber\\
|L|^2 + |K|^2 = |L_1|^2 + |K_1|^2 = 1.
\end{eqnarray}

The condition  (\ref{Eqn51}) according to which the potential doesn't depend on the momentum for the excitation
with the energy $E(\vec P)$, is fulfilled if the coefficients are related as

\begin{eqnarray}
\label{Eqn52a} L_1 = - K; \qquad K_1 = L.
\end{eqnarray}

These conditions are also consistent with  Equations (\ref{Eqn50}) for  wave functions.

This means that the "physical" electron moves in such a way that its states are transformed in the phase space
of the orthogonal wave functions (\ref{Eqn28a}),(\ref{Eqn31b}),(\ref{Eqn33b}),(\ref{Eqn33ñ})but the amplitudes
of its "internal" charge distributions are not changed. These results, however, are valid only for the case of
neglecting the interaction with the transverse electromagnetic field.

Substituting the superpositions  (\ref{Eqn52}) to Equations (\ref{Eqn50}) we use the following relations:

\begin{eqnarray}
\label{Eqn53} (\vec \alpha \vec P) \Psi(\vec r) = \left( \begin{array}{c}
i f (r) (\vec \sigma \vec P)\Omega_{1/2, 1, M}\\
g(r) (\vec \sigma \vec P) \Omega_{1/2,0,M}
\end{array} \right); \quad
(\vec \alpha \vec P) \Psi^c(\vec r) = \left( \begin{array}{c}
g_1 (r) (\vec \sigma \vec P)\Omega_{1/2, 1, M}\\
-i f_1(r) (\vec \sigma \vec P) \Omega_{1/2,0,M}
\end{array} \right);\nonumber\\
(- i \vec \alpha \vec \nabla + \beta m_0 + e_0 \varphi) \Psi(\vec r) = E(0) \left( \begin{array}{c}
g (r) \Omega_{1/2, 0, M}\\
i f(r) \Omega_{1/2,1,M}
\end{array} \right); \nonumber\\
(-i \vec \alpha \vec \nabla + \beta m_0 + e_0 \varphi)  \Psi^c(\vec r) = - E(0) \left( \begin{array}{c}
-i f_1 (r) \Omega_{1/2, 0, M}\\
g_1(r) \Omega_{1/2,1,M}
\end{array} \right),
\end{eqnarray}

and similar formulas for the functions  $\tilde{\Psi}(\vec r); \tilde{\Psi^c}(\vec r)$;  $\sigma_i$ are the
Pauli matrixes.

For Equations  (\ref{Eqn50}) to be fulfilled for any vector $\vec r$ and it is necessary to set the coefficients
of spherical spinors equal to the same indexes  $l$. The corresponding radial functions are proved to be the
same under these conditions and the following system of equations for the spinors $\chi_{0,1}^{\pm}$ is obtained

\begin{eqnarray}
\label{Eqn54} i L (\vec \sigma \vec P)\chi^+_1 + K ( E + E_0)\chi_1^+ = 0; \quad i K (\vec \sigma \vec P)\chi^+_0
+ L(
E - E_0)\chi_0^+ = 0;\nonumber\\
L (\vec \sigma \vec P)\chi^+_0 + i K ( E + E_0)\chi_0^+ = 0; \quad  K (\vec \sigma \vec P)\chi^+_1 + i L(
E - E_0)\chi_1^+ = 0;\nonumber\\
L_1 (\vec \sigma \vec P)\chi^-_1 - i K_1 ( E - E_0)\chi_1^- = 0; \quad  K_1 (\vec \sigma \vec P)\chi^-_0 - i L_1(
E + E_0)\chi_0^- = 0;\nonumber\\
i L_1 (\vec \sigma \vec P)\chi^-_1 -  K_1 ( E - E_0)\chi_1^- = 0; \quad i K_1 (\vec \sigma \vec P)\chi^-_0 -L_1(
E + E_0)\chi_0^- = 0.
\end{eqnarray}

Spin variables in Equations (\ref{Eqn54}) are also separated. In order to show it one can use, for example, the
relation between the coefficients resulting from the 4th equation in  (\ref{Eqn54}),

$$
\chi_1^+ = i \frac{K (\vec \sigma \vec P)\chi^+_1}{L( E - E_0)}
$$

in the first one of these equations. As the result there exists a non-trivial solution of these equations for
two branches of the energy spectrum

\begin{eqnarray}
\label{Eqn55} E_{e,p} = \pm \sqrt{E^2_0 + P^2},
\end{eqnarray}

referring to the electron and positron zones, respectively (Fig.1). The same expressions can be obtained for all
conjugated pairs of the equations in (\ref{Eqn54}). The coefficients in the wave functions (\ref{Eqn52}) can be
found on the account of the normalization condition:

\begin{eqnarray}
\label{Eqn56} L^e = K^e_1 = \frac{P}{\sqrt{P^2 + (E_e - E_0)^2}}; \quad K^e = - L^e_1 =
\frac{E_e - E_0}{\sqrt{P^2 + (E_e - E_0)^2}}; \nonumber\\
L^p = K^p_1 = \frac{P}{\sqrt{P^2 + (E_e + E_0)^2}}; \quad K^p = - L^p_1 = -\frac{E_e + E_0}{\sqrt{P^2 + (E_e  +
E_0)^2}}.
\end{eqnarray}

One can see that this set of coefficients coincides with the set of spinor components for solving the Dirac
equation for a free electron with the observed mass $ m = E_0$. Thus, the results of this section show that the
"internal" structure of the "physical" electron (positron) considered in this paper is consistent with the
experimental energy dispersion (\ref{Eqn47}) for a real free particle due to the relativistic invariance of the
Dirac equation.

However, in order to make the interpretation of the wave packet  (\ref{Eqn47a}) as the state vector for a
"physical" particle complete we should consider one more problem that has not been discussed yet. The matter is
that the exact vectors of the one-particle states of EPF should not only be normalized according to
(\ref{Eqn47b}) but be also orthogonal

\begin{eqnarray}
\label{Eqn56a} < \Phi_1 (\vec P_1) |  \Phi_1 (\vec P)> = \delta (\vec P - \vec P_1).
\end{eqnarray}

It is evident that the approximate representation (\ref{Eqn47a}) for the one-particle state vector doesn't
satisfy this condition . The reason for non-orthogonality is  that the variational principle (\ref{Eqn50})used
for determining the energy spectrum of quasi-particle excitations allows to satisfy the exact law of the
momentum conservation only on the average as in Equation (\ref{Eqn47b}).  At the same time the orthogonality
condition  (\ref{Eqn56a}) holds true for the exact eigenfunction of the momentum operator.

The rigorous method of taking into account the translational symmetry in the theory of strong coupling for the
"polaron" problem was elaborated in the works of Bogolubov \cite{Bogol} and Gross \cite{Gross}. Let's remember
that this method was based on the introduction of the collective variable  $\vec R$ conjugated to the total
momentum operator  $ \hat {\vec P}$, the canonical character of the transformation caused by three new variables
$R_i$ being provided by the same number of additional conditions imposed on the other variables of the system.
In the "polaron" problem the quantum field interacting with the particle contributes to the total momentum of
the system. It allows to impose these conditions on the canonical field variables \cite{Bogol}, \cite{Gross} and
the concrete form of the variable transformation is based essentially on the permutation relation for the
operator of the boson field .

The considered problem has some specific features in comparison with the "polaron" problem. Firstly, the
formation of the one-particle wave packet is the many-particle effect because this packet includes all initial
states of EPF as the fermionic field. Secondly, its self-localization is provided by the polarization potential
of the scalar field which doesn't contribute to the total momentum of the system. Therefore we use another
approach in order to select the collective coordinate  $\vec R$. Let us return to the configuration
representation in the hamiltonian  (\ref{Eqn18}) , where QED is considered to be the totality of N $( N
\rightarrow \infty)$ point electrons interacting with the quantum EMF in the Coulomb gauge ò\cite{Heitler}:

\begin{eqnarray}
\label{10} \hat H = \sum_{a=1}^{N} \{  \vec{\alpha}_a [\hat { \vec{p}_a} + e_0 \hat{ \vec{ A} } (\vec{r}_a)] +
\beta_a m_0 + e_0 \hat \varphi (\vec{r}_a)\} - \nonumber\\  \frac{1}{2} \int d \vec{r}( \vec{\nabla}\hat \varphi
(\vec{r}))^2 + \sum_{\vec k \lambda} \omega (\vec k) \hat n_{\vec k \lambda};
\nonumber\\
\omega (\vec k) = k; \quad \hat n_{\vec k \lambda} = c^+_{\vec k \lambda}c_{\vec k \lambda}; \quad \lambda = 1,2;
\nonumber\\
\hat{ \vec{ A} } (\vec{r}) = \sum_{\vec k \lambda} \frac{1}{\sqrt{2 k \Omega}} \vec {e}^{(\lambda)} [ c_{\vec k
\lambda} e^{i \vec k \vec r} + c^+_{\vec k \lambda} e^{-i \vec k \vec r}]
\end{eqnarray}

Here $\Omega$ is the normalized volume; $c^+_{\vec k \lambda}(c_{\vec k \lambda})$ are the operators of the
creation (annihilation) of quanta of the transversal electromagnetic field, the quantum having the wave vector
$\vec k$, polarization  $\vec {e}^{(\lambda)}$ and energy $\omega (\vec k) = k$. The sign of the interaction
operators differs from the standard one because the parameter $e_0$ is introduced as having a positive value.

As it was stated above the zero approximation of nonperturbative QED is defined only by a strong interaction of
electrons with the scalar field, the interaction with the transversal field is taken into account in the
framework the standard perturbation theory when the conservation of the total momentum is provided automatically
\cite{Akhiezer}. Therefore, while describing the quasi-particle excitation we should consider the conservation
of the total momentum only for the system of electrons. In the considered representation it is defined by the
sum of the momentum operators of individual particles

\begin{eqnarray}
\label{11} \hat {\vec P} = \sum_{a=1}^{N} \hat {\vec{p}_a};  \nonumber\\
\hat {\vec P} |  \Phi_1 (\vec P)> = \vec P |  \Phi_1 (\vec P)>.
\end{eqnarray}

It means that in the configuration space the variable $\vec R$ conjugated to the total momentum is simply a
coordinate of the center of mass and the desired transformation to new variables is as follows:

\begin{eqnarray}
\label{12}\vec r_a = \vec R + \vec{\rho}_a; \quad  \vec R = \frac{1}{N}\sum_{a=1}^{N}\vec r_a; \quad
\sum_{a=1}^{N} \vec{\rho}_a = 0; \nonumber\\
\hat {\vec{p}_a} = - i \vec {\nabla}_a = - \frac{i}{N}\vec {\nabla}_R + \hat {\vec{p}'_a}; \quad \hat {\vec P} =
-i\vec {\nabla}_R; \quad \hat {\vec{p}'_a} =- i \vec {\nabla}_{\rho_a} + \frac{i}{N}\sum_{b=1}^{N}\vec
{\nabla}_{\rho_b}; \quad \sum_{a=1}^{N} \hat {\vec{p}'_a} = 0.
\end{eqnarray}

The hamiltonian (\ref{10}) with new variables has the following form

\begin{eqnarray}
\label{13} \hat H = \sum_{a=1}^{N} \{  \vec{\alpha}_a [- \frac{i}{N}\vec {\nabla}_R + \hat {\vec{p}'_a}
 + e_0 \hat{ \vec{ A} } (\vec{\rho}_a + \vec R)] + \beta_a m_0 + e_0 \hat \varphi (\vec{\rho}_a + \vec R)\}
 - \nonumber\\
\frac{1}{2} \int d \vec{r}( \vec{\nabla}\hat \varphi (\vec{r}))^2 + \sum_{\vec k \lambda} \omega (\vec k) \hat
n_{\vec k \lambda}.
\end{eqnarray}

It is essential to note that the matrix elements of an arbitrary operator in a new configuration representation
should be calculated in accordance with the following norm (we introduce a special notation for that norm)

\begin{eqnarray}
\label{14} << \Phi_1 | \hat M | \Phi_2 >> = \int d \vec R \prod_{a} d \vec {\rho}_a \Phi_1^* (\vec R, \{\vec
{\rho}_a\})\hat M \Phi_2 (\vec R, \{\vec {\rho}_a\}).
\end{eqnarray}

Let us denote by $\hat H_0$ that part of the operator (\ref{13}) which doesn't depend on the transversal EMF and
describes the internal structure of "physical" particles . Actually the operator  $\hat H_0$ doesn't depend  on
the coordinate $\vec R$ either, because of its commutativity with the operator of the total momentum of the
system of electrons. This also follows from the well known result  \cite{Heitler} that in the Coulomb gauge the
scalar potential could be excluded from the hamiltonian. As the result the operator $\hat H_0$ depends only on
the vector differences  $(\vec {r}_a - \vec {r}_b) = (\vec {\rho}_a - \vec {\rho}_b)$ not changing with the
simultaneous translation of all the coordinates. As the consequence, the eigenfunctions of the hamiltonian $\hat
H_0$ depend on the coordinate $\vec R$ in the same way as for a free particle:

\begin{eqnarray}
\label{15}  \Phi (\vec R, \{\vec {\rho}_a\}) = \frac{1}{(2\pi)^{3/2}}e^{i\vec P\vec R} |\Phi_1 (\vec
P, \{\vec {\rho}_a\})>;\nonumber\\
\hat H_0 \rightarrow \hat H_0 (\vec P) = \sum_{a=1}^{N} \{  \vec{\alpha}_a [ \frac{1}{N}\vec {P} + \hat
{\vec{p}'_a} ] + \beta_a m_0 + e_0 \hat \varphi (\vec{\rho}_a )\} -  \frac{1}{2} \int d \vec{r}(
\vec{\nabla}\hat \varphi (\vec{r}))^2
\end{eqnarray}

Further calculations consist in returning to the field representation by the variables   $ \vec {\rho}_a$ in the
limit $( N \rightarrow \infty)$ and in using the approximate trial wave packet \ref{Eqn20}) for the state vector
$|\Phi_1 (\vec P)>$ . As the result one obtains the functional for the energy  $E(\vec P)$ which coincides
completely with the expression (\ref{Eqn49}). It should be noted that the averaging of the operator $\hat H_0
(\vec P)$ in the secondary quantization was calculated using the relation which follows from the well known
density of states when EPF is quantized in the normalized volume  $\Omega$ \cite{Akhiezer}

$$
\lim_{N \to \infty} \frac{1}{N} \sum_{\vec p} = 1.
$$

Thus, the orthogonalized and normalized set of states for the EPF one-particle excitations corresponding to the
"physical" electrons (positrons) is defined as follows

\begin{eqnarray}
\label{16} |\Phi^{(0)}_1(\vec P) > \simeq \frac{1}{(2\pi)^{3/2}}e^{i\vec P\vec R} \int d \vec{q} \{
U_{\vec{q}s}(\vec P) a^+_{\vec{q} s} + V_{\vec{q}s}(\vec P) b^+_{\vec{q} s} \} | 0; 0;\varphi(\vec r)>;
\end{eqnarray}

with the norm (\ref{14}) and the coefficient functions  $U_{\vec{q}s}(\vec P);V_{\vec{q}s}(\vec P)$ which are
combined by the Fourier transformation  $(\ref{Eqn26})$ of the wave functions $\Psi (\vec r, \vec P); \Psi^c
(\vec r, \vec P)$ defined by the formulas  (\ref{Eqn52}).

\section{Perturbation theory for QED with a large value of the initial coupling constant }

It is well known that the renormalizability is one of the most important features of QED and it is confirmed by
the coincidence of its results with the experimental data.  Actually it means that the calculated
characteristics of the real electromagnetic processes don't depend on the initial parameters  $e_0, m_0$ but
only on the observed values of e and m. So, we should clearly show that the effects of the nonperturbative QED
refer to the internal structure of the "physical" electron (positron) but its interaction with a transversal
electromagnetic field is defined by the standard perturbation theory with the fine structure constant $\alpha
\ll 1$. As for a scalar field, this requirement is met by the normalized condition (\ref{Eqn21a}).

One more problem is to demonstrate that the integrals which define the corrections to the zero approximation of
the nonperturbative QED are converged without introducing any additional regularizating parameters. This
condition was one of the reason for the present work.

To solve these problems  the form of the perturbation theory which uses the basis of states corresponding to
"physical" electron (positron) should be considered. Taking into account the results of the last section let us
represent the QED hamiltonian  (\ref{Eqn18}) in the following form :

\begin{eqnarray}
\label{Eqn57} \hat H = \hat H_0 + \hat H_I; \nonumber\\
\hat H_0 = \int d \vec{r}\{ : \hat \psi^* (\vec{r}) [ - i\vec \alpha ( \vec {\nabla}_{\vec R} + \vec
{\nabla}_{\vec r}) + \beta m_0] \hat \psi (\vec{r}): + e_0 \hat \varphi (\vec{r}) :\hat \rho (\vec{r}): -
\frac{1}{2} ( \vec{\nabla}\hat \varphi
(\vec{r}))^2\} + \sum_{\vec k \lambda} \omega (\vec k) \hat n_{\vec k \lambda};\nonumber\\
\hat H_I = e_0 \int d \vec{r} : \hat \psi^* (\vec{r}) [\vec \alpha \hat{ \vec{ A} } (\vec{r} + \vec{R})] \hat
\psi (\vec{r}):.
\end{eqnarray}

The field operators  $\hat \psi(\vec{r})$ and the density operator $\hat \rho (\vec{r})$ are defined by
expressions (\ref{Eqn18}),(\ref{Eqn19}) and we  the symbol $: \hat A :$ of the normal ordering is used to
exclude the vacuum energy (\ref{Eqn24}) in the EPF excitation spectrum. The fact that the hamiltonian depends on
the coordinate $\vec R$ canonically conjugated to the total EPF momentum determines the local violation of
symmetry of the system. This situation was discussed by Bogolubov  \cite{quasi} for a lot of concrete physical
systems. However, the global symmetry of the system    \cite{quasi} is reconstructed when the observed values
are averaged by the coordinate  $\vec R$ in accordance with the norm (\ref{14}).

 As distinct from the canonical perturbation theory we can't find exactly the total spectrum of  eigenvectors
of the operator  $\hat H_0$ for the considered case of the electron strong coupling with a scalar field . In the
previous sections only the spectrum of the one-particle excitations was found approximately in the main order of
the expansion in terms of $\alpha_0^{-1}$ power. In particular, even the calculation of the two-particle
excitation spectrum requires the analys of the system analogous to the "bipolaron" problem in the Pekar model
\cite{bipolaron}. Such analysis is out of the framework of the present paper. Therefore we consider only such
matrix elements of the perturbation operator   $H_I$ which are due to the transitions between the states
corresponding to the free "physical" electron (positron) and an arbitrary number of quanta of the transversal
electromagnetic field. Let us write the obvious form of these states using definition (\ref{16}) for the
one-particle wave packet

\begin{eqnarray}
\label{Eqn58} |\Phi_1(\vec P), {N_{\vec k,\lambda}} > \simeq \frac{1}{(2\pi)^{3/2}}e^{i\vec P\vec R}  \int d
\vec{q} \{ U_{\vec{q},s}(\vec P) a^+_{\vec{q} s} + V_{\vec{q},s}(\vec P) b^+_{\vec{q} s} \}{ \frac{(c^+_{\vec k,
\lambda})^{N_{\vec k,\lambda}}}{\sqrt{N_{\vec k,\lambda}!}}} |0; 0;\varphi(\vec r)>.
\end{eqnarray}

In order to be definite, let us calculate the matrix element of the operator   $\hat H_I$ for the transition of
the "physical" electron between the states with the momenta  $\vec P$ and $\vec P_1$ together with emitting one
quantum of the transversal EMF. The norm (\ref{14}) is used when calculating  this element

\begin{eqnarray}
\label{Eqn59}   \Gamma_{\lambda} (\vec P, \vec k) = <<\Phi_1(\vec P_1), \vec k,\lambda |\hat H_I |\Phi_1(\vec
P), 0
>> =
\delta (\vec P - \vec P_1 - \vec K) \gamma_{\lambda} (\vec P, \vec k); \nonumber\\
 \gamma (\vec P, \vec k)\simeq  \frac{e_0}{\sqrt{2 k \Omega}}\sum_{s,s'} \sum_{\mu,\nu}\int d \vec{r}\int
 \frac{ d \vec{q}}{(2\pi)^{3/2}}
\int \frac{ d \vec{q}'}{(2\pi)^{3/2}}  \{ U^*_{\vec{q}',s'}(\vec{P}_1) U_{\vec{q},s}(\vec{P})u^*_{\vec{q}'s'\mu}
(\vec {\alpha} \vec {e}^{\lambda})_{\mu,\nu}u_{\vec{q}s\nu} - \nonumber\\
V^*_{\vec{q}',s'}(\vec{P}_1) V_{\vec{q},s}(\vec{P})v_{-\vec{q}'s'\mu} (\vec {\alpha} \vec
{e}^{\lambda})_{\mu,\nu}v^*_{-\vec{q}s\nu}\} e^{i(\vec q - \vec {q}' - \vec k) \vec r}.
\end{eqnarray}

In order to calculate the renormalized vertex function $\gamma (\vec P, \vec k)$ let us find the under integral
expression through the spinor wave functions of one-particle states in the coordinate representation by means of
the relations  (\ref{Eqn48})

\begin{eqnarray}
\label{Eqn60}
 \gamma_{\lambda} (\vec P, \vec k) =  \frac{e_0}{\sqrt{2 k \Omega}}\int d \vec{r}\{ \Psi^*  (\vec r, \vec {P}_1)
(\vec {\alpha} \vec {e}^{\lambda})\Psi  (\vec r, \vec {P})
 - \Psi^{c,*}  (\vec r, \vec {P}_1)
(\vec {\alpha} \vec {e}^{\lambda})\Psi^c  (\vec r, \vec {P}) \} e^{- i \vec k \vec r}; \nonumber\\
\vec {P}_1= \vec {P} - \vec {k}.
\end{eqnarray}

Let's substitute the linear combinations  (\ref{Eqn52}) in this expression and use the analytical form of the
functions $\Psi, \Psi^c, \tilde{\Psi}, \tilde{\Psi}^c$ for the resting "physical" electron given in Eqs.
(\ref{Eqn28a}),(\ref{Eqn31b}),(\ref{Eqn33b}),(\ref{Eqn33ñ}). Then the transition to the two-component spinors
should be written as follows:

\begin{eqnarray}
\label{Eqn61}
 \gamma_{\lambda}(\vec P, \vec k) =  i\frac{e_0}{\sqrt{2 k \Omega}}\int d \vec{r}\{ [g^2(r)(\chi^{+*}_0 (\vec {\sigma}
 \vec {e}^{\lambda})\chi^+_0) - f^2(r) (\chi^{+*}_1 (\vec {\sigma}
 \vec {e}^{\lambda})\chi^+_1)][K^* (\vec {P}_1) L (\vec {P}) - \nonumber\\ L^* (\vec {P}_1) K (\vec {P})]
-[g_1^2(r)(\chi^{-*}_1 (\vec {\sigma}
 \vec {e}^{\lambda})\chi^-_1) - \nonumber\\ f_1^2(r) (\chi^{-*}_0 (\vec {\sigma}
 \vec {e}^{\lambda})\chi^-_0)] [K_1^* (\vec {P}_1) L_1 (\vec {P}) - L_1^* (\vec {P}_1) K_1 (\vec {P})]\}
 e^{- i \vec k \vec r}.
\end{eqnarray}

In accordance with the results of Section 3 the radial wave functions of the internal structure of the
"physical" electron are localized in the domain   $r \leq (b m_0)^{-1}$. It means that if the vertex function is
calculated with the photon momentum less than a certain finite but large enough value

\begin{eqnarray}
\label{Eqn62} k < k_0 \simeq b m_0 \simeq 328 m \simeq 167 (MeV),
\end{eqnarray}

the exponent in the integrals (\ref{Eqn61}) can be substituted by unit  $\exp (- i \vec k \vec r) \simeq 1$.
Besides, the spinor and angular variables can be excluded by means of Eq.  (\ref{Eqn52}) for the spinor wave
functions describing the internal state of the "physical" electron. Thus, for example,

$$
K (\vec {P})(\vec {\sigma} \vec {e}^{\lambda})\chi^+_0 = \frac{iL (\vec {P})}{E + E_0}(\vec {\sigma} \vec
{P})(\vec {\sigma} \vec {e}^{\lambda})\chi^+_0=\frac{iL (\vec {P})}{E + E_0}(\vec {P} \vec
{e}^{\lambda})\chi^+_0.
$$

Taking into account the normalization of the spinors  $\chi^{+,-}_{0,1}$ one can find that the vertex function
has the following form when the condition (\ref{Eqn62}) is fulfilled

\begin{eqnarray}
\label{Eqn63}
 \gamma_{\lambda}
  (\vec P, \vec k) =  \frac{e_0}{\sqrt{2 k \Omega}}
 L (\vec {P})L (\vec {P}_1)[\frac{(\vec {P} \vec
{e}^{\lambda})}{E(\vec P) + E_0} - \frac{(\vec {P}_1 \vec {e}^{\lambda})}{E(\vec {P}_1) + E_0}]\nonumber\\
\int d
\vec{r}\{
[g^2(r)+ f^2(r)] - [g_1^2(r)+ f_1^2(r)]\} = \nonumber\\
= \frac{e}{\sqrt{2 k \Omega}}
 L (\vec {P})L (\vec {P}_1)[\frac{(\vec {P} \vec
{e}^{\lambda})}{E(\vec P) + m} - \frac{(\vec {P}_1 \vec {e}^{\lambda})}{E(\vec {P}_1) + m}].
\end{eqnarray}

We have also used the normalization conditions for the radial functions and the relations (\ref{Eqn37}) which
define the observed charge e and mass m  of the "physical" electron. The obtained expression for the vertex
function completely coincides with the result of the standard perturbation theory
 \cite{Heitler}. This result is not obvious a priori and it seems to be a serious argument in favour
of the nonperturbative QED. It also shows that after the renormaliation our approach leads mainly to the same
results for the observed characteristics of the electromagnetic processes as the standard form of the
perturbation theory in QED.

And finally, the last question we are are going to consider briefly in the present paper is concerned with the
regularization of the diverged integrals in QED. It is evident from  expression (\ref{Eqn61}) that the behavior
of the renormalized vertex function $\gamma (\vec P, \vec k)$ at large k is defined by the Fourier-image of the
radial functions connected with the form-factor of the charge distribution in the "physical" electron. If the
variational solutions  (\ref{Eqn39}) are used, then the asymptotic behavior of the vertex function at large $k
\gg k_0$ is defined by the following dependence

\begin{eqnarray}
\label{Eqn64}
 \gamma (\vec P, \vec k)\quad \sim \quad  \frac{e}{\sqrt{2 k \Omega}}\frac{k_0^4}{(k^2 + 4 k_0^2)^2},
\end{eqnarray}

and converges rather quickly to zero .

Thus, the cut off momentum $k_{max} \sim k_0$ appears naturally in the nonperturbative QED and it has quite a
definite physical meaning. It is well known (for example, \cite{Akhiezer}) that the introduction of such
momentum provides the regulariztion of all integrals in QED.  However, it should be stressed that the wave
functions describing the charge distribution in the "physical" electron are not completely spherically
symmetrical . Therefore, the effective cut off of the integration domain in the integrals over the momenta could
be not spherically symmetrical either unlike of the usual methods of regularization. But these corrections
should be rather small because of a quick decrease of the functions under the integrals at the domain boundary
in accordance with   (\ref{Eqn64}). The concrete estimation of these corrections is connected with rather
cumbersome calculations and is out of the framework of the present work.

\section{Conclusions}

Thus, a new logically consistent solution of quantum electrodynamics equations is constructed in the present
paper. It is based on the supposition that the existing form of QED is a closed and self-consistent theory of
the interacting electron-positron and electromagnetic fields.

The main advantage of the obtained solution is the possibility to connect in the finite form the observed and
initial parameters of QED without introducing any additional parameters. This solution, of course, doesn't
exclude the interaction of EPF and EMF with other quantum fields, but demonstrates the possibility of the finite
renormalization in QED itself without taking into account these fields and other external factors.

According to our results the non-renormalized form of QED corresponds to the theory of strong interaction
between electron-positron and scalar electromagnetic fields. The one-particle excitation of EPF leads to a local
modification of the vacuum states of the system, with quite a strong classical self-consistent field appearing
as the result. In the field of this potential the "physical" electron (positron) is formed as the quasi-particle
excitation consisting of  two strongly coupled charge distributions of  opposite signs. The effective masses of
both distributions are large. But the integral charges of both distributions are almost equal to each other and
their binding energy is comparable with the sum of their masses. As the result of the compensation of these
large values the observed values for the charge and mass of the "physical" electron (positron) appear. It is
shown that the electromagnetic processes for the "physical" electron-positron field are defined by the
renormalized charge and mass of the electron. But the internal structure of the "physical" electron defines the
unobserved cut off momentum in the vertex function. It is also shown that the states and energy spectrum of the
"physical" particles satisfy  the conditions of the relativistic invariance.

Certainly, the calculations given in our paper are approximate and should be developed in various directions. In
particular, it is important to formulate the nonpertubative QED in a covariant form on the basis of the Lorentz
gauge, to find a decisive evidence that the results of nonperturbative and standard  QED are equivalent after
the renormalization, to investigate the effect of other fields on the obtained solution, to consider the
spectrum of two-particle excitations. The formulation of the nonperturbative QED on the basis of the path
integrals is of special interest. It is impossible to analyze all these problems in the framework of one paper
but it seems to us that the  physical picture of the renormalzation considered here could be the foundation for
further discussions and investigations in the above mentioned directions.

\section{Acknowledgements}

The author is grateful to  $\begin{tabular}{|r|} \hline
L.I.Komarov\\
\hline
\end{tabular}$ for discussions and to A.M.Zaitzeva, M.N.Polozov, A.P.Ulyanenkov and Burg\"azi ,
International Scientific and Technical Center (Grant B-626) for the
support of this work .

\newpage
\begin{center}
{CAPTION FOR FIGURES\\
\bigskip

FIG.1

\bigskip

Scheme of the one-particle levels for the vacuum ($E^{(0)}(\vec P)$) and "physical" ($E(\vec P)$)
  electron-positron field.}

\end{center}


\begin{thebibliography}{}

\bibitem{Dirac} P.A.M.Dirac, {\it The Principles of Quantum Mechanics},
(Oxford, 1958).

\bibitem{Feynman} R.P.Feynman, Nobel Lecture {\it Science},
{\bf 153},(1966), 699.

\bibitem{Mors} P.M.Morse and H.Feshbach, {\it Methods of
Theoretical Physics}, (N.Y.:McGraw-Hill Book Co., 1953).

\bibitem{electric} W.Heisenberg and H.Euler
 {\it Ztschr. Phys.}, {\bf 98}, (1936), 714-732.

\bibitem{kulon} A.B.Migdal, {\it Fermions and Bosons in the Strong
Fields}, (Moscow, Nauka, 1978).

\bibitem{Zimann} J.M.Ziman, {\it Principles of the Theory
of Solids}, (Cambridge, The University Press, 1972).

\bibitem{Akhiezer} A.I.Akhiezer and V.B.Beresteckii, {\it Quantum
Electrodynamics}, (Moscow, Nauka, 1969).

\bibitem{Bogoliubov} N.N.Bogoliubov and D.V.Shirkov, {\it Introduction to the
Theory of Quantum Fields}, (Moscow, Nauka, 1973).

\bibitem{Scharf95} G. Scharf, {\it Finite Quantum Electodynamics:
the Causal Approach}, (USA, John Wiley and Sons, Inc., 2001).

\bibitem{Scharf01} G. Scharf, {\it Quantum Gauge Theories:
a True Ghost story}, (Berlin, Heidelberg, New York, Springer Verlag, 1995).

\bibitem{Kleinert} Proceedings: {\it Fluctuating Paths and Fields}, (Singapore,
World Scientific, 2001).

\bibitem{Lifshitz} E.M.Lifshitz and L.P.Pitaevskii, {\it Relativistic
Quantum Theory. Part II}, (Moscow, Nauka, 1971).

\bibitem{Blinder} S.M.Blinder, {\it Eur. J. Phys.},
{\bf 24},(2003), 271-275.

\bibitem{Heitler} W.Heitler, {\it The Quantum Theory of Radiation},
(Oxford, The Clarendon Press, 1954).

\bibitem{Pekar} S.I.Pekar, {\it Zh. Exper. Teor. Fiz.},
{\bf 16},(1946), 769-774.



\bibitem{Caswell} W.E.Caswell, {\it Ann. Phys. (N.Y.)},
{\bf 123},(1979), 153.

\bibitem{OM82} I.D.Feranchuk and L.I.Komarov, {\it Phys. Lett. A},
{\bf 88},(1982), 211-213.

\bibitem{OM95} I.D.Feranchuk, L.I.Komarov, I.V.Nichipor and
A.P.Ulyanenkov, {\it Ann. Phys. (N.Y.)}, {\bf 238},(1995), 370-440.

\bibitem{OM96} I.D.Feranchuk, L.I.Komarov and
A.P.Ulyanenkov, {\it J. of Phys. A: Math. and General}, {\bf 29},(1996), 4035-4047.

\bibitem{Acta} I.D.Feranchuk, L.I.Komarov,
A.P.Ulyanenkov et al,{\it Acta Cryst. A}, {\bf 58},(2002), 370-384.

\bibitem{quasi} N.N.Bogoliubov,  {\it Quasi-mean Values in The Problems
of Statistical Mechanics. In "Selected Works", v.3}, (Kiev, "Navukova Dumka", 1971).

\bibitem{Bogol} N.N.Bogoliubov,  {\it Uspekhi Matematicheskih Nauk},
{\bf 2},(1950), 3-24.

\bibitem{Tyablikov} S.V.Tyablikov, {\it Zh. Exper. Teor. Fiz.},
{\bf 21},(1951), 377-388.

\bibitem{Rochev} V.E.Rochev, {\it J. of Phys. A: Math. and General}, {\bf 33},(2000), 7379-7406 .

\bibitem{Gross} Gross E.P. {\it Annals of Physics (NY)},
{\bf 99},(1976), 1.

\bibitem{Feynmanpol} R.P.Feynman, {\it Phys. Rev.},
{\bf 97},(1955), 660.

\bibitem{Fradkin} E.S.Fradkin, {\it Proceedings of Fiz. Inst.
of Soviet Academy of Science}, {\bf 29},(1965), 1-154; {\it Nucl. Phys.} {\bf 76}, (1966), 588.

\bibitem{Pekar1} S.I.Pekar {\it Zh. Exper. Teor. Fiz.},
{\bf 16},(1946), 335-339.

\bibitem{Krylov} L.I.Komarov, E.V.Krylov and I.D.Feranchuk,
{\it Zh. Numerical Math. and Math. Fiz.}, {\bf 18},(1978), 681-691.

\bibitem{Landau} L.D.Landau and E.M.Lifshitz {\it Quantum Mechanics}, (Moscow, Nauka, 1963).


\bibitem{impuls} G. H\"oler,{\it Zh. Phyzik},{\bf 146},(1956), 372.


\bibitem{PV} L.D.Landau and S.I.Pekar,{\it Zh.of Exper. and Theoret. Physics},{\bf 18},(1948), 419.

\bibitem{bipolaron} V.L.Vinetzkii,{\it Zh.of Exper. and Theoret. Physics},{\bf 40},(1961), 1459.

\end{thebibliography}
\end{document}